\documentclass{aa}  

\usepackage{graphicx}
\usepackage{txfonts}
\usepackage{lipsum}
\usepackage{natbib}
\usepackage{scalerel}
\usepackage{siunitx}
\usepackage{subcaption}         
\usepackage{lscape}             
\usepackage{placeins}           
\usepackage[pdfencoding=auto,psdextra]{hyperref}
\hypersetup{
    colorlinks=true,
    linkcolor=blue,
    filecolor=magenta,      
    urlcolor=blue,
    citecolor=blue
}

\begin{document}

   \title{Optical Counterparts of MeerKLASS L-band and UHF-band surveys}


%
\author{
  M.~Klein$^{1}$\thanks{\email{matthias.klein@physik.lmu.de}} \and S.~Mangla$^{1,2}$, J.~Mohr$^{1}$ \and S.~Chatterjee$^{3}$ \and K.~Grainge$^{4}$, S. Paul$^{4}$ \and M.~G.~Santos$^{3,5}$ \and O.~M.~Smirnov$^{6,5,7,8,9}$ \and C.~Tasse$^{10,6}$ \and L.~Wolz$^{4}$}

\institute{University Observatory, LMU Faculty of Physics, Scheinerstr. 1, 81679 Munich, Germany 
\and Excellence Cluster ORIGINS, Boltzmannstrasse 2, D-85748 Garching, Germany
\and Department of Physics \& Astronomy, University of the Western Cape, Cape Town 7535, South Africa
\and Jodrell Bank Centre for Astrophysics, Department of Physics \& Astronomy, The University of Manchester, Manchester M13 9PL, UK
\and South African Radio Astronomy Observatory (SARAO), 2 Fir Street, Cape Town, 7925, South Africa
\and Centre for Radio Astronomy Techniques and Technologies (RATT), Department of Physics and Electronics, Rhodes University, Makhanda, 6140, South Africa
\and Institute for Radioastronomy, National Institute of Astrophysics (INAF IRA), Via Gobetti 101, 40129 Bologna, Italy
\and Astrophysics, Department of Physics, University of Oxford, Keble Road, Oxford, OX1 3RH, UK
\and Breakthrough Listen, Astrophysics, Department of Physics, University of Oxford, Keble Road, Oxford, OX1 3RH, UK
\and GEPI \& ORN, Observatoire de Paris, Université PSL, CNRS, 5 Place Jules Janssen, 92190 Meudon, France}

   \date{Received July, 20XX}

 
  \abstract
  {Wide-area radio continuum surveys require reliable optical and infrared counterpart identification, but high optical source densities and extended or multi-component radio morphologies make this challenging.}
   {We present optical and infrared counterpart catalogs for MeerKLASS L-band and UHF-band on-the-fly continuum sources using KiDS DR5 and DESI Legacy Imaging Surveys DR10 (LS DR10), including counterpart probabilities, redshifts, and host-galaxy properties.}
   {We developed the Stellar-mass Enhanced Density Association (SEDA) method, an empirical framework that compares candidate densities around radio positions with those in a position-displaced control catalog. For galaxies we use positional offset, stellar mass, and redshift; quasar candidates are treated separately using offset and mid-infrared selection. A second-pass search associates multi-component radio sources with common hosts.}
    {In the L-band survey, we identify 20,400 KiDS-based counterparts with $P_{\mathrm{true}}>0.5$, corresponding to 66\% of L-band sources within the KiDS footprint. In the UHF-band survey, we identify 61,633 LS DR10 counterparts, corresponding to 81\% of the radio sources. Spectroscopic redshifts are available for 22\% and 44\% of the L-band and UHF-band counterparts, respectively. The redshift distributions show low-redshift star-forming galaxies, intermediate-redshift radio galaxies, and a high-redshift tail dominated by quasars. SEDA also recovers rare radio-loud quasars, including QSO J2318-3113 at $z=6.44$ and UHF\_DR1\,J+111111.8+053626.6 at $z=5.24$.}
   {SEDA provides a data-driven route to counterpart identification for wide-area radio surveys with complex source morphologies. The catalogs enable future MeerKLASS studies of radio source populations, host-galaxy demographics, and rare high-redshift radio quasars.}


   \maketitle
 \nolinenumbers

\begin{figure*}
\begin{center}
\centering
\includegraphics[width=1\linewidth]{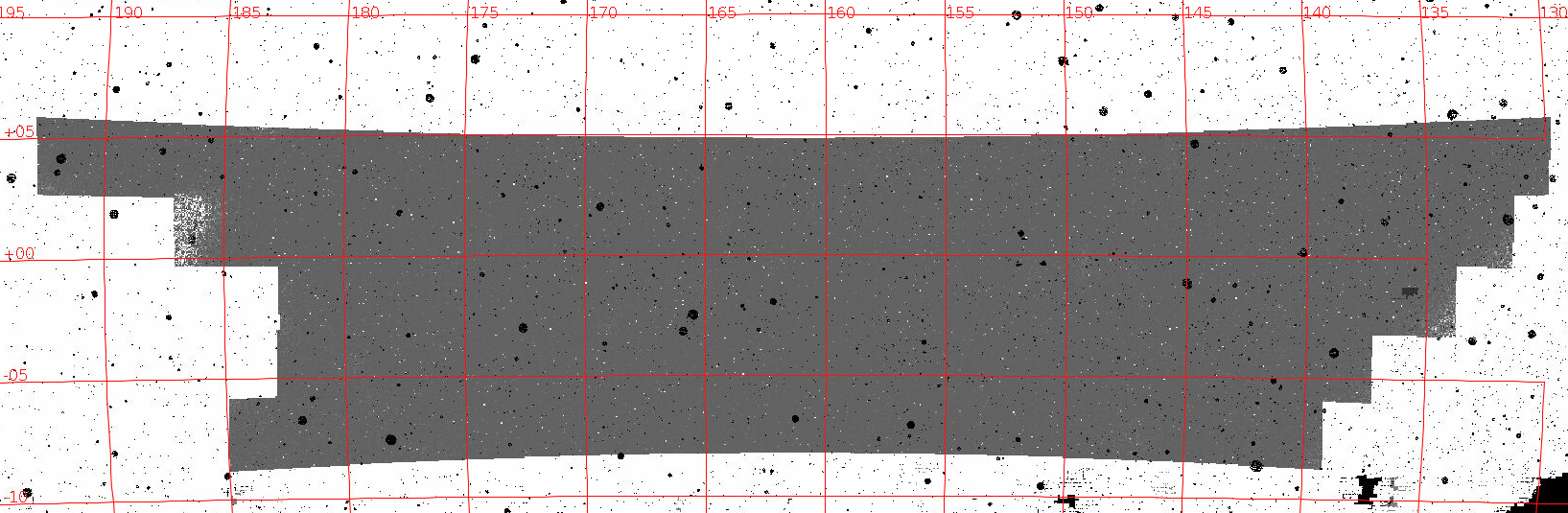}
\includegraphics[width=1\linewidth]{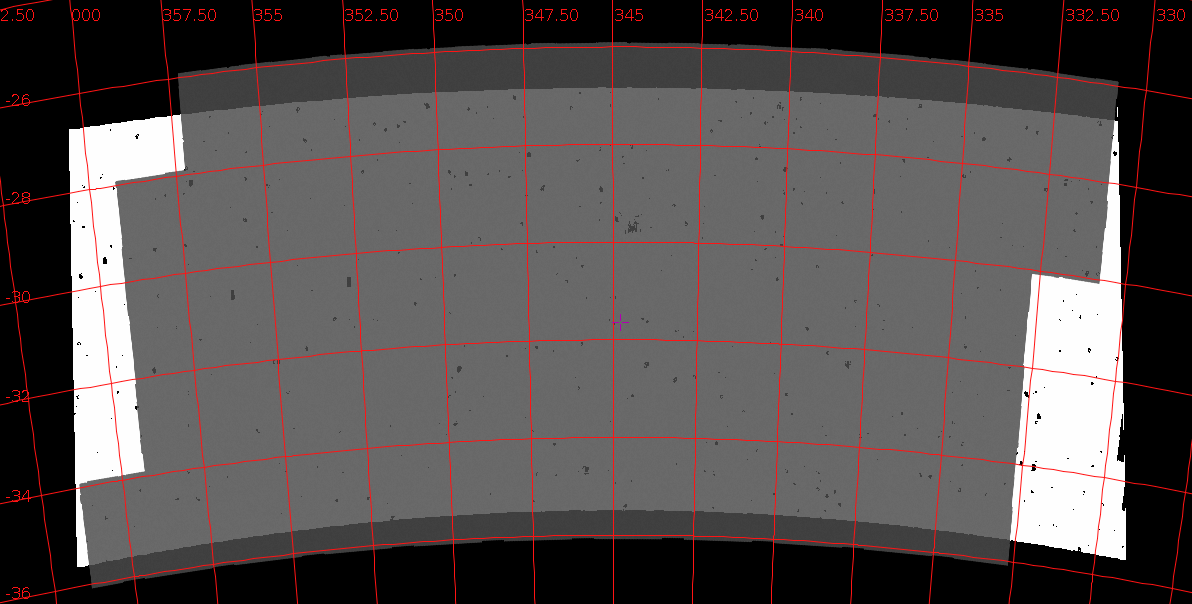}
\caption{Optical coverage of the MeerKLASS surveys. Top: MeerKLASS UHF-band radio image (grey) overlaid on the LS DR10 coverage map. Bottom: MeerKLASS L-band image overlaid on the KiDS DR5 coverage map.}
\label{fig:coverage}
\end{center}
\vspace{-0.3cm}
\end{figure*}

\section{Introduction}
Wide-area radio continuum surveys have become a cornerstone of modern extragalactic astronomy, providing an unobscured view of star formation and black-hole accretion across cosmic time. Synchrotron emission from star-forming galaxies (SFGs) and active galactic nuclei (AGN) offers a dust-insensitive tracer of these processes and enables the construction of statistically robust samples from the local Universe to high redshift. Over the past decade, radio surveys with SKA precursor and pathfinder facilities have greatly expanded the accessible parameter space in depth, angular resolution, and sky coverage. Building on earlier large-area surveys such as FIRST \citep{Becker1995FIRST} and NVSS \citep{Condon1998NVSS}, current surveys including RACS \citep{Hale2021RACS}, LoTSS \citep{Shimwell2019LoTSS}, VLASS \citep{Lacy2020VLASS}, ASKAP/EMU \citep{Hopkins2025EMU}, MIGHTEE \citep{Heywood2022MIGHTEE}, and VLA-COSMOS \citep{Smolcic2017VLACOSMOS} now probe radio source populations over a wide range of flux densities, angular scales, and cosmic volumes.

The MeerKAT Large Area Synoptic Survey (MeerKLASS; \citealt{Santos2017}) represents a major step forward in this context. Conducted with the MeerKAT interferometer, MeerKLASS combines wide sky coverage, high sensitivity, and sub-arcminute angular resolution through a fast, constant-elevation scanning strategy originally designed for \ion{H}{i} intensity-mapping cosmology. Using an on-the-fly (OTF) interferometric imaging approach, these commensal observations deliver deep, contiguous radio continuum maps over hundreds to thousands of square degrees \citep{Chatterjee2025}. In its L-band and UHF-band components, MeerKLASS occupies a unique niche between shallow all-sky surveys such as NVSS and RACS, and deep pencil-beam surveys such as MIGHTEE and VLA-COSMOS, providing an optimal dataset for statistical studies of radio-selected galaxy populations.

The first MeerKLASS continuum data releases demonstrate the high fidelity of the OTF imaging technique and the good astrometric and photometric performance of the resulting radio catalogs \citep{Paul2025,Mangla2025}. The MeerKLASS L-band survey reaches rms sensitivities of a few tens of $\mu$Jy\,beam$^{-1}$ at angular resolutions of order $\sim10$--$30''$, enabling the detection of large samples of star-forming galaxies out to moderate redshifts and radio AGN across a broad range of luminosities. At the same time, the wide survey footprint ensures extensive overlap with current and forthcoming optical and infrared surveys, including KiDS, the DESI Legacy Imaging Surveys, 4MOST, \emph{Euclid}, and the Rubin Observatory Legacy Survey of Space and Time (LSST). This combination makes MeerKLASS particularly well suited for large-scale optical and infrared counterpart studies.

The scientific exploitation of these radio datasets critically depends on the identification and characterization of their host galaxies at optical and infrared wavelengths. Radio continuum observations alone provide limited information on redshift, stellar mass, or star-formation properties, and the interpretation of radio luminosities and morphologies requires complementary multi-wavelength data. Optical and near-infrared follow-up is therefore essential to transform radio source catalogs into physically meaningful galaxy samples, enabling the separation of AGN- and star-formation-dominated systems, the measurement of radio luminosity functions, and the investigation of environmental and evolutionary trends.

Counterpart identification is straightforward only for compact radio sources with accurate positions and relatively sparse optical backgrounds. In deeper optical and infrared surveys, chance associations become increasingly common, but enable higher true associations to radio sources. Extended and multi-component radio sources can show intrinsic offsets between the cataloged radio position and the true host galaxy, further complicating the counterpart search. Classical nearest-neighbor matching therefore becomes insufficient for a significant subset of sources. Statistical approaches, most prominently likelihood-ratio methods \citep{Sutherland1992}, have been widely used to incorporate positional uncertainties and magnitude-dependent background densities, and modern radio surveys often combine such techniques with visual inspection or morphology-aware classification \citep[e.g.][]{McAlpine2012,Williams2019}. These developments highlight the need for association methods that are both statistical and flexible enough to handle the heterogeneous radio morphologies encountered in wide-area surveys.

In this paper, we present the optical and infrared follow-up and host-galaxy identification of radio sources detected in the MeerKLASS L-band and UHF-band surveys. We combine the MeerKLASS radio catalogs with KiDS DR5 \citep{Wright2024} and DESI Legacy Imaging Surveys DR10 (LS DR10) \citep{Legacysurveys19} data, augmented by available spectroscopic redshifts, to construct multi-wavelength catalogs of radio-selected galaxies. Our analysis focuses on the identification methodology, the statistical properties of the matched sample, and the resulting redshift and host-galaxy distributions. Particular emphasis is placed on assessing identification reliability and on characterizing the optical properties of the radio source population across different source classes.

This work provides a step toward the full scientific exploitation of the MeerKLASS radio data and establishes a foundation for future studies of galaxy evolution using upcoming, deeper data releases. By linking the MeerKLASS radio surveys to optical and infrared datasets over large sky areas, we enable a wide range of investigations into the nature and evolution of star-forming galaxies and AGN in the era leading up to the Square Kilometre Array.


\section{Data}
In the subsections below we describe the parent MeerKLASS DR1 catalogs in L- and UHF-bands.  Moreover, we describe the multiband optical, NIR and mid IR surveys used in the cross-matching analysis of the MeerKLASS DR1 sample.

\subsection{MeerKLASS radio data}
Our radio source samples are drawn from two components of the MeerKAT Large Area Synoptic Survey (MeerKLASS), a commensal radio-continuum and \ion{H}{i} intensity-mapping program carried out with MeerKAT using fast constant-elevation scans and on-the-fly (OTF) interferometric imaging \citep{Santos2017,Chatterjee2025}. In this observing mode, interferometric visibilities are recorded continuously during the scan observations and processed into wide-area continuum images and source catalogs. This on-the-fly interferometric imaging provides an efficient combination of survey speed, sensitivity, and sky coverage while maintaining high image fidelity over large contiguous areas \citep{Chatterjee2025}.

\subsubsection{MeerKLASS UHF-band survey}\label{sec:UHF}

For the lower-frequency sample, we use sources from the first data release (DR1) of the MeerKLASS UHF-band OTF continuum survey \citep{Paul2025}. These observations were carried out with MeerKAT in the 544--1088\,MHz frequency range using fast constant-elevation scanning and were processed with a dedicated OTF imaging pipeline. The DR1 dataset is based on approximately 12\,hr of early-science observations and covers an area of $\sim800\,\mathrm{deg}^2$ within the DESI imaging footprint.

The resulting continuum images are centered at a frequency of $\sim816$\,MHz and reach an rms sensitivity of $\sim35\,\mu$Jy\,beam$^{-1}$ in the deepest regions, with a typical angular resolution of $\sim32'' \times 17''$. The associated source catalog contains 75,823 radio sources above a signal-to-noise ratio of 7. The catalog has been validated through cross-matching with external radio surveys, demonstrating sub-arcsecond astrometric accuracy and a robust flux-density scale \citep{Paul2025}. The large survey area and direct overlap with wide-field optical imaging make this dataset particularly well suited for statistical identification of optical and infrared counterparts.

\subsubsection{MeerKLASS L-band survey}\label{sec:Lband}

For the higher-frequency sample, we use sources from the first data release of the MeerKLASS L-band OTF continuum survey \citep{Mangla2025}. These observations were obtained in the 856--1712\,MHz frequency range using the same fast-scanning OTF strategy. The current release is based on eight observing blocks, corresponding to approximately 13.5\,hr of usable data and covering an area of $\sim268\,\mathrm{deg}^2$.

The L-band images reach a median rms sensitivity of $\sim33\,\mu$Jy\,beam$^{-1}$ and have a median angular resolution of approximately $25.5'' \times 7.8''$. The corresponding source catalog contains 34,874 radio sources. The footprint of the L-band survey overlaps with deep optical imaging from KiDS, enabling high-quality counterpart identification for a subset of the radio sample \citep{Mangla2025}.

\subsection{Optical and near-infrared imaging data}

To identify optical counterparts to the MeerKLASS radio sources and to derive photometric properties of the host galaxies, we use wide-area public optical and near-infrared imaging surveys matched to the respective MeerKLASS footprints. For the UHF-band survey we rely on the DESI Legacy Imaging Surveys Data Release 10 (hereafter LS DR10), while for the L-band survey we additionally exploit the deeper KiDS DR5 data where available. 

\subsubsection{DESI Legacy Imaging Surveys DR10}\label{sec:lsdr10}

The DESI Legacy Imaging Surveys provide a homogeneous, wide-area optical imaging dataset designed for target selection of the Dark Energy Spectroscopic Instrument (DESI) and for a broad range of extragalactic studies \citep{Legacysurveys19}. The surveys combine imaging from multiple facilities, including DECam, Bok, and Mayall/MOSAIC, together with mid-infrared data from the \textit{WISE} satellite \citep{WISE}. The full Legacy Surveys footprint covers approximately 25,000\,deg$^2$ of extragalactic sky.

In this work, we use LS DR10. Both MeerKLASS radio surveys lie in sky regions covered by DECam imaging, with available $g$, $r$, $i$, and $z$-band data. However, the region covered by the MeerKLASS L-band survey does not show homogeneous coverage in all bands, making LS DR10 less suitable as the main optical counterpart catalog for that field. In contrast, the MeerKLASS UHF-band footprint is well covered by DECam imaging. The relevant point-source depths are approximately $\sim24.8$, $\sim24.2$, and $\sim23.3$\,mag in the $g$, $r$, and $z$ bands, respectively.

Photometry and source detection in LS DR10 are based on forward modeling of sources using \texttt{The Tractor} \citep{TractorLang16}, enabling consistent photometry and deblending across the full survey footprint. Forced photometry is performed at the positions of detected sources across all bands, ensuring uniform multi-band measurements. We further make use of photometric redshifts described in \citet{Zhou21,Zhou23}.

The combination of large sky coverage ($>10^4\,\mathrm{deg}^2$), homogeneous processing, and the inclusion of mid-infrared photometry makes LS DR10 particularly well suited for wide-area cross-identification of radio sources, especially in current and upcoming MeerKLASS UHF-band fields, which lie almost entirely within the DESI footprint.

\subsubsection{KiDS DR5}\label{sec:Kids}

For the MeerKLASS L-band footprint, we additionally use data from the fifth data release of the Kilo-Degree Survey (KiDS DR5), which provides deep optical imaging with excellent image quality \citep{Wright2024}. KiDS DR5 covers a total area of 1,347\,deg$^2$ in the $u$, $g$, $r$, and $i$ bands, with particular emphasis on high image quality for weak-lensing applications. Most importantly for this work, KiDS DR5 provides homogeneous coverage of $\sim87\%$ of the L-band survey, making it the primary resource for the search for counterparts to MeerKLASS L-band radio sources.

The survey reaches a median $r$-band 5$\sigma$ depth of $\sim24.8$\,mag and achieves a median seeing of $\sim0.7''$. In addition, KiDS is complemented by near-infrared imaging from the VIKING survey, providing matched photometry in the $Z$, $Y$, $J$, $H$, and $K_s$ bands.

The combination of deep optical imaging, high spatial resolution, and multi-wavelength coverage makes KiDS DR5 particularly valuable for identifying faint counterparts to radio sources and for deriving accurate photometric redshifts. This is especially important for the MeerKLASS L-band sample, where the smaller survey area allows the use of deeper ancillary data for detailed source characterization.

We augment the main KiDS DR5 catalog, designed to provide reliable estimates for the bulk of optical sources, with the KiDS DR4 bright-galaxy catalog \citep{Bilicki2021} and the KiDS DR4 quasar catalog \citep{Nakoneczny2021}. We further add WISE \citep{WISE} photometry to the catalog by cross matching KiDS sources to the latest version of the unWISE catalog \citep{UnWISE} and the LS DR10 catalogs. This enables a consistent WISE-based quasar selection in both radio surveys. These catalogs provide improved redshift estimates and source properties for their respective target populations. Their addition is motivated by the expectation that the radio-selected sample contains a disproportionately large contribution from low-redshift star-forming galaxies and high-redshift radio quasars, whereas the bulk of optically selected galaxies lies at intermediate redshift.

\begin{figure}
\begin{center}
\centering
\includegraphics[width=1\linewidth]{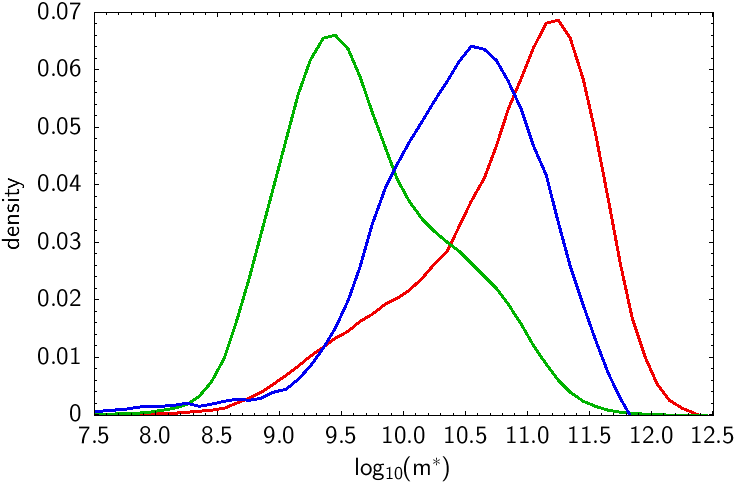}
\caption{Stellar mass distribution of sources with small ($<2''$) radio--optical offsets (red) and of local field galaxies (green) with offsets of ($40''-60''$). A subset of sources with small offsets at low redshifts ($z<0.2$) is shown in blue.}
\label{fig:stellmass}
\end{center}
\vspace{-0.3cm}
\end{figure}

\section{Method}

Identifying optical counterparts of radio sources is a non-trivial task, particularly for modern wide-area surveys that combine high optical source densities with complex radio morphologies. The simplest approach is nearest-neighbor positional matching within a fixed search radius. This method is computationally efficient and performs well for compact radio sources with high positional accuracy, but it becomes increasingly unreliable in deep optical surveys because of the high probability of chance alignments.

More sophisticated statistical methods have therefore been developed. In particular, the likelihood-ratio (LR) technique \citep{Sutherland1992} provides a probabilistic framework that accounts for positional uncertainties, the magnitude distribution of true counterparts, and the background source density. This approach, together with related probabilistic cross-identification methods, has become a standard tool for radio--optical counterpart identification in modern surveys \citep[e.g.][]{Smith2011,McAlpine2012,Williams2019}.

Additional complexity arises from the extended and multi-component nature of many radio sources, where the radio emission may be offset from the host galaxy. Recent approaches therefore combine statistical matching techniques with morphological information and, in some cases, visual or machine-learning-based classification to improve identification reliability \citep[e.g.][]{Williams2019,Hardcastle2023}.

Motivated by the characteristics of the MeerKLASS data, we adopt a data-driven approach that empirically measures the offset distribution of candidate hosts around radio source positions and compares it to the background distribution estimated from a position-displaced control catalog within the same survey footprint.
This approach naturally captures the combined effects of positional uncertainties and intrinsic offsets between radio emission and host galaxies, which can arise from extended or multi-component radio sources. Furthermore, we incorporate physical 
characteristics of the host candidates, such as stellar mass and redshift, and treat different source populations separately. In this sense, our method can be viewed as a generalized likelihood-based approach that replaces 
source positional uncertainties and host magnitude distributions with empirically extracted distributions in radio-optical offset, stellar mass and redshift that are directly calibrated with the survey data. In the following, we refer to this approach as the Stellar-mass Enhanced Density Association (SEDA) method.

\subsection{The Stellar-mass Enhanced Density Association (SEDA) method}
Extragalactic radio sources can broadly be associated with three main host populations: quasars, radio galaxies, and star-forming galaxies. While quasars and radio galaxies both belong to the class of active galactic nuclei (AGN), they differ significantly in their appearance in optical and infrared data. Star-forming galaxies are detected in radio through star-formation-driven synchrotron emission and typically dominate at lower radio luminosities, while radio emission from AGN is produced by synchrotron emission from the central core as well as the jets and lobes.

The probability of a galaxy being radio loud is strongly dependent on stellar mass \citep[e.g.,][]{Best2005}. For star-forming galaxies, radio luminosity is also expected to correlate with stellar mass through the star-forming main sequence. In contrast, stellar mass is neither a reliable indicator of radio luminosity for quasars nor straightforward to measure, because the AGN can contribute substantially to the total optical and infrared flux of the system. Quasars are, however, rare compared to the general galaxy population in optical and infrared catalogs and can be selected efficiently using mid-infrared colors from \textit{WISE}. We therefore split the counterpart search into two sub-populations: quasars and a stellar-mass-dependent population consisting of star-forming galaxies and radio galaxies. For quasar candidates, we primarily use the radio-to-optical offset to estimate whether a candidate is a true match to a given MeerKLASS source.

The radio luminosity function and the stellar mass function of star-forming galaxies differ, and the mixture of radio-emitting populations is therefore expected to change with redshift. At low redshift, the increasing sensitivity to low-luminosity radio sources makes lower-mass star-forming galaxies increasingly important. Conversely, at high redshift, star-forming galaxies are less likely to be detected at the flux limits of the MeerKLASS catalogs and are therefore less likely to be suitable host-galaxy candidates. For the stellar-mass-sensitive population, we therefore use radio-to-optical offset, stellar mass, and redshift simultaneously to estimate the likelihood of a given counterpart being the true host of the MeerKLASS source.

\begin{figure}
\begin{center}
\centering
\includegraphics[width=1\linewidth]{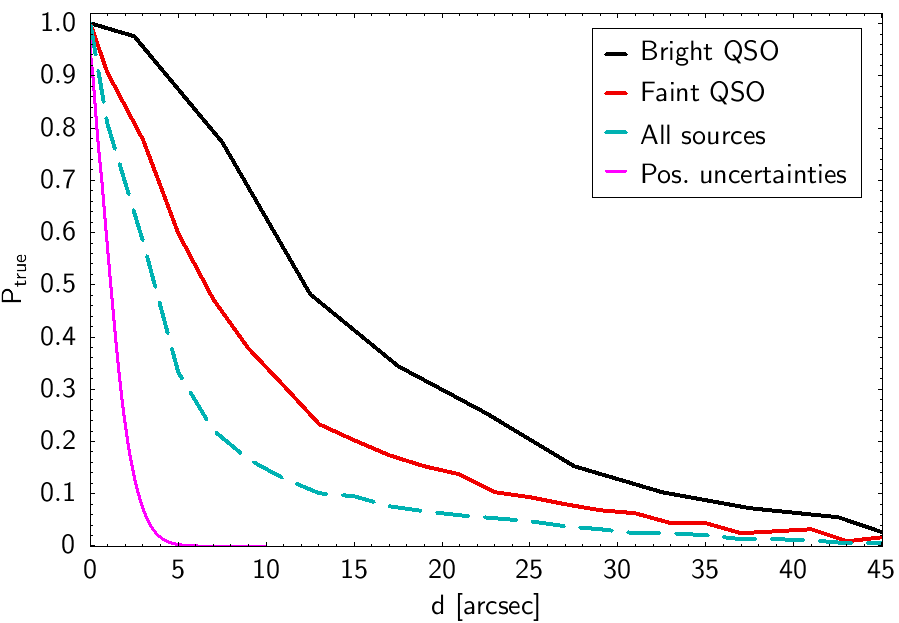}
\caption{Empirical probability estimate $P_{\mathrm{true}}$ as a function of offset for bright and faint quasar candidates (black and red lines), the full dataset (cyan dashed line), and the expectation from positional uncertainties and source density (magenta).}
\label{fig:petrueoffset_qso}
\end{center}
\vspace{-0.3cm}
\end{figure}

\subsubsection{Probability $P_{\mathrm{true}}$ of candidate being the host galaxy}
The empirical probability estimate of a candidate being the true host depends on four properties: stellar mass, radio-to-optical offset, redshift, and whether the source is classified as a quasar. For the non-quasar population, we evaluate the probability estimate $P_{\mathrm{true}}(d,M_\star,z)$ that a source is the true optical counterpart, given its offset $d$, stellar mass $M_\star$, and redshift $z$.

To do so, we first collect all optical and infrared sources within a one-arcminute radius around each MeerKLASS source. This radius is much larger than the typical formal positional uncertainty of MeerKLASS sources, which is usually at the level of one arcsecond or below, and allows for physically driven offsets between the radio emission and the host galaxy. The PSFs of the MeerKLASS DR1 surveys are highly elliptical, with axis ratios of $\sim1/2$ and $\sim1/3$ for the UHF-band and L-band surveys, respectively. Even in the presence of physically driven offsets, we expect the ellipticity of the MeerKLASS beams to affect the symmetry of the offset distribution. To account for this, we use elliptical distances based on the axis ratios given in Sects.~\ref{sec:UHF} and \ref{sec:Lband}.

We measure the source density of optical and infrared sources, $n_\mathrm{S}(d,M_\star,z)$, around MeerKLASS sources in 30 radial bins out to $45''$, 52 stellar-mass bins between $7.3<\log_{10}{\left(M_\star/M_\odot\right)}<12.5$, and three redshift bins. We also measure the source density in the same stellar-mass and redshift bins in an outer annulus from $45''$ to $60''$, which serves as an estimate of the background source density, $n_\mathrm{BG}$. The empirical probability estimate for a galaxy being the true host is then
\begin{equation}
    P_{\mathrm{true}}(d,M_\star,z)= 1 - \frac{n_\mathrm{BG}(M_\star,z)}{n_\mathrm{S}(d,M_\star,z)}.
\end{equation}
Given the size of the current dataset, we split the data into three redshift slices: $z<0.2$, $0.2\leq z<0.4$, and $z\geq0.4$. The highest-redshift bin is expected to be largely free of star-forming galaxies, while the lower-redshift bins are expected to be helpful in capturing the increased sensitivity to lower-mass star-forming galaxies.  With the larger size and depth of future MeerKLASS datasets, we expect to be able to improve the treatment of the redshift dependence of the counterpart probability.

\begin{figure}
\begin{center}
\centering
\includegraphics[width=1\linewidth]{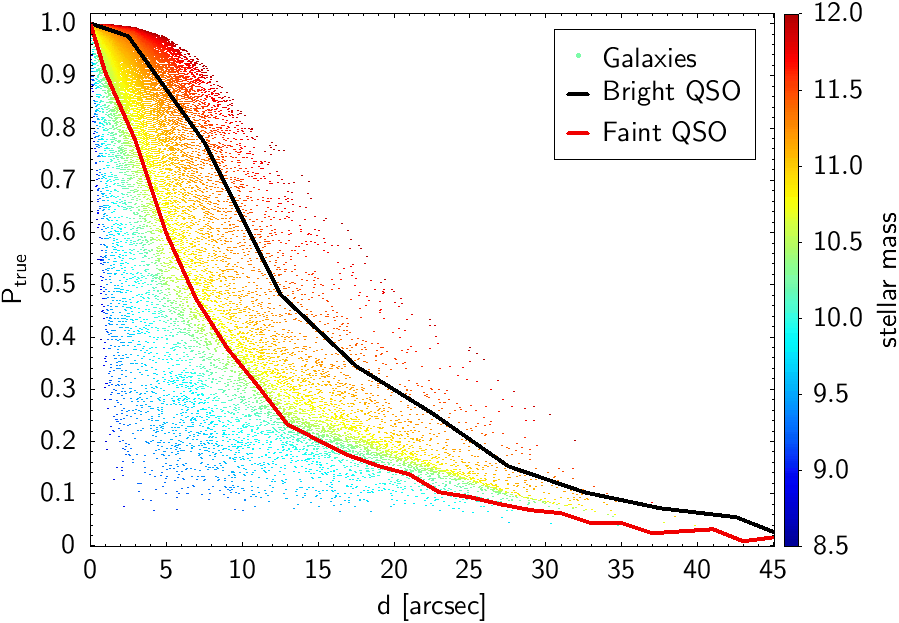}
\caption{Empirical probability estimate $P_{\mathrm{true}}$ as a function of positional offset for candidates with $z>0.4$. The color coding highlights the stellar-mass dependence. The corresponding relations for bright and faint quasar candidates (see Fig.~\ref{fig:petrueoffset_qso}) are shown as black and red lines for comparison.}
\label{fig:petrueoffset_stellmass}
\end{center}
\vspace{-0.3cm}
\end{figure}

\subsubsection{Treatment of quasars}\label{sec:Quasars}
The stellar-mass estimates for quasars are not reliable and are therefore not informative for the counterpart probability estimate. At the same time, quasars are relatively rare sources and occupy a specific region of color space. We select quasar candidates as point-like sources with $w1-w2>-0.2$. This relatively relaxed WISE color cut is motivated by the fact that the search is performed locally around radio positions rather than as a blind field selection. As a result, the prior probability of a source being physically associated with a radio emitter is already significantly enhanced compared to the general field population. The color selection is sufficient to strongly suppress the contribution from stars and passive galaxies, while the point-source selection further suppresses the contribution from star-forming galaxies.

To further reduce contamination from non-quasar populations, we impose a magnitude limit of $w2<21$. This cut reduces contamination caused by large scatter in $w1-w2$ colors and decreases the chance-match rate by excluding faint candidates. Using nearby matches ($d<2''$), we find that this magnitude cut retains $>92\%$ of all WISE-based quasar candidates. Because the number density of quasar candidates depends strongly on flux, we further split the sample into bright and faint quasar candidates at $w2=19.5$. Analogous to the definition for star-forming galaxies and radio galaxies, we define the empirical probability estimate of a quasar candidate being the true host as
\begin{equation}
    P_{\mathrm{true}}(d)= 1 - \frac{n_\mathrm{BG}}{n_\mathrm{S}(d)}.
\end{equation}
The probability estimate depends smoothly on offset and on whether the source falls into the bright or faint quasar subset. In future improvements of SEDA, when we have larger samples available, we will model this dependence more continuously by measuring the source density simultaneously in offset and $w2$ magnitude.

In Fig.~\ref{fig:petrueoffset_qso}, we show $P_{\mathrm{true}}(d)$ for faint and bright quasar candidates as well as for all sources irrespective of source type. The value of $P_{\mathrm{true}}(d)$ remains significantly higher for bright quasars at all offsets. Faint quasars also show a higher $P_{\mathrm{true}}(d)$ than the full source sample. This is mainly driven by the suppression of the background population of unlikely true hosts, while the offset distributions are approximately similar in all three cases.

\begin{figure}
\begin{center}
\centering
\includegraphics[width=1\linewidth]{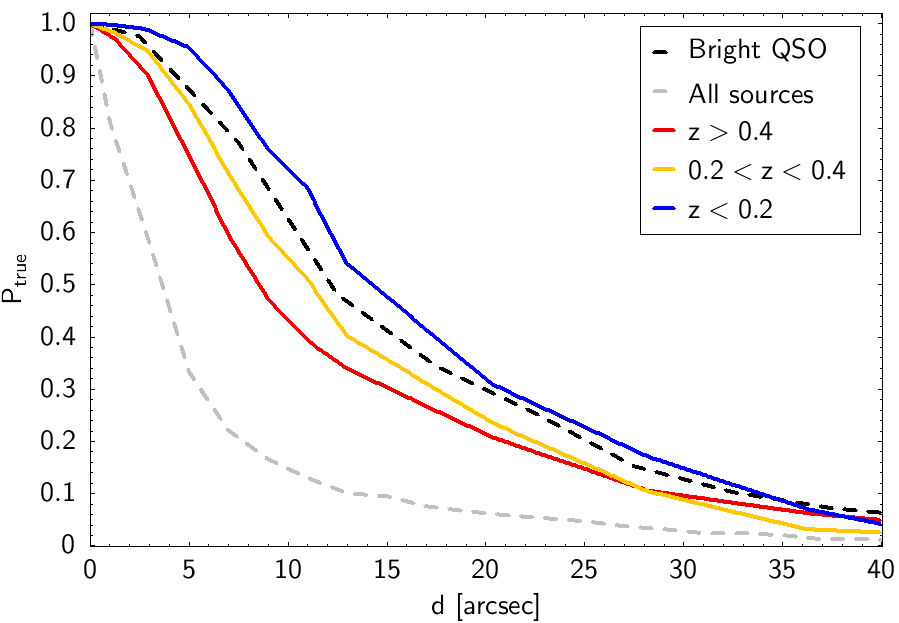}
\caption{Empirical probability estimate $P_{\mathrm{true}}$ for a $M_\star=10^{11}\,\mathrm{M}_{\odot}$ source as a function of positional offset for the three redshift bins. The $P_{\mathrm{true}}(d)$ relations for bright quasar candidates and all sources are shown as black and grey dashed lines for comparison.}
\label{fig:petrueoffset_redshift}
\end{center}
\vspace{-0.3cm}
\end{figure}

\subsubsection{Radial dependence of $P_{\mathrm{true}}(d,M_\star,z)$}

In Fig.~\ref{fig:petrueoffset_stellmass}, we show the radial trend of $P_{\mathrm{true}}$ color-coded by stellar mass. For high-mass sources, $P_{\mathrm{true}}$ remains high even at larger positional offsets. This results mainly from two effects: massive galaxies are less abundant and therefore less likely to occur as chance matches, and high-mass galaxies are typical hosts of radio galaxies, as shown in Fig.~\ref{fig:stellmass}. Towards lower stellar masses, the background density increases while the fraction of true hosts at a given stellar mass decreases.

In Fig.~\ref{fig:petrueoffset_redshift}, we show the radial trend of $P_{\mathrm{true}}$ for the three redshift bins at a fixed stellar mass of $M_\star=10^{11}\,\mathrm{M}_\odot$. At fixed stellar mass, $P_{\mathrm{true}}$ is, on average, highest for the lowest-redshift bin and lowest for the highest-redshift bin. Several effects contribute to this behavior. At low redshift, radio sources tend to be more spatially resolved, broadening the offset distribution of true hosts. In addition, the background density is reduced, and the stellar-mass distributions of true hosts and field galaxies are shifted to lower stellar masses (see Fig.~\ref{fig:stellmass}), as more low-luminosity star-forming galaxies exceed the detection threshold of the survey.

\begin{figure}
\begin{center}
\centering
\includegraphics[width=1\linewidth]{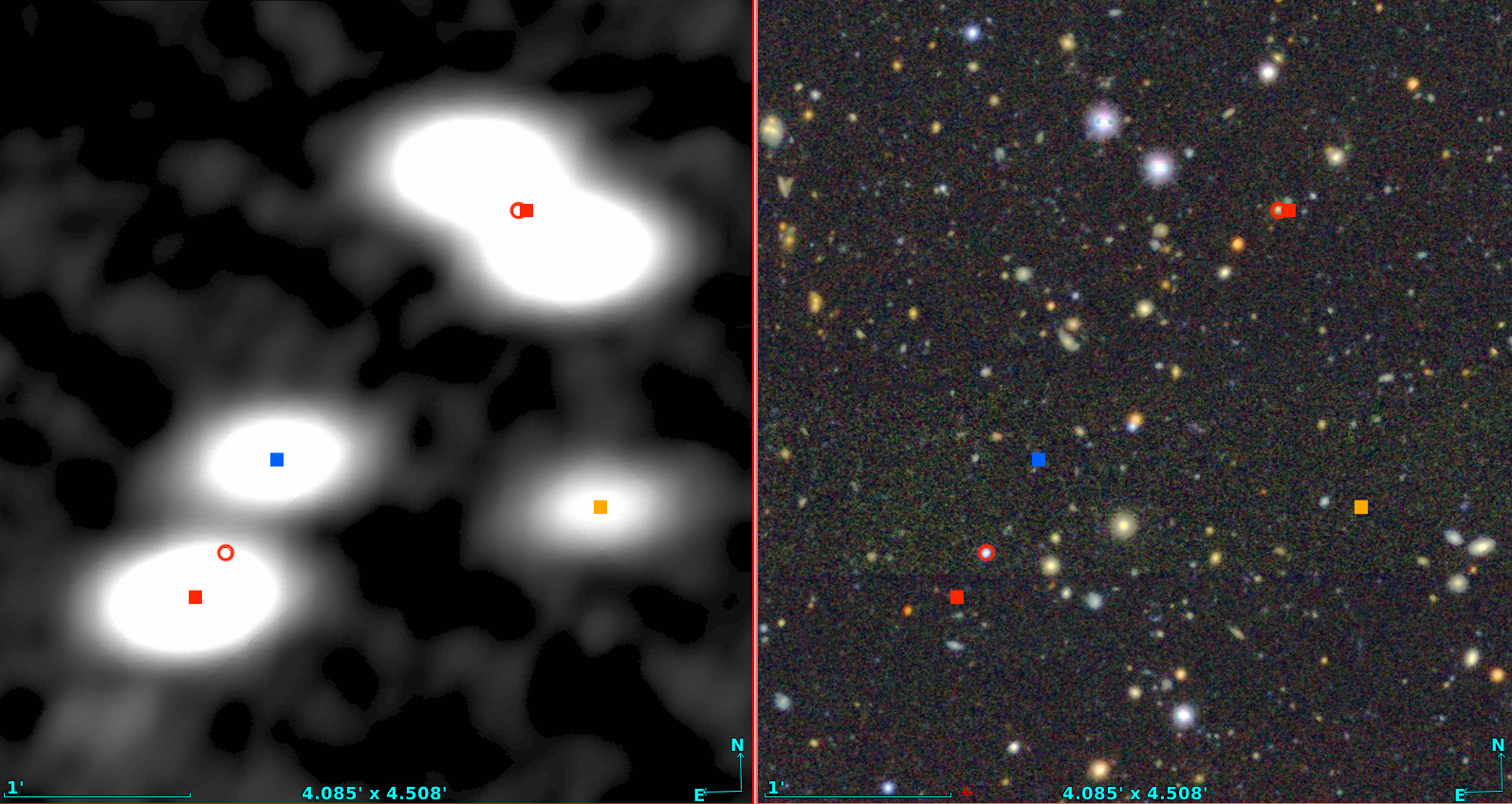}
\caption{Counterpart search and source merging. Left: MeerKLASS UHF-band image of a $4'\times4'$ region near MeerKLASS-UHF\_DR1\,J+124001.6+034037.5. Right: LS DR10 $gri$ color-composite image of the same region. Individual radio detections are marked as filled squares. The color coding indicates the primary source (red), other components (blue), and sources without confirmed counterparts (yellow). Optical counterparts are shown as circles.}
\label{fig:merging}
\end{center}
\vspace{-0.3cm}
\end{figure}

\begin{figure}
\begin{center}
\centering
\includegraphics[width=1\linewidth]{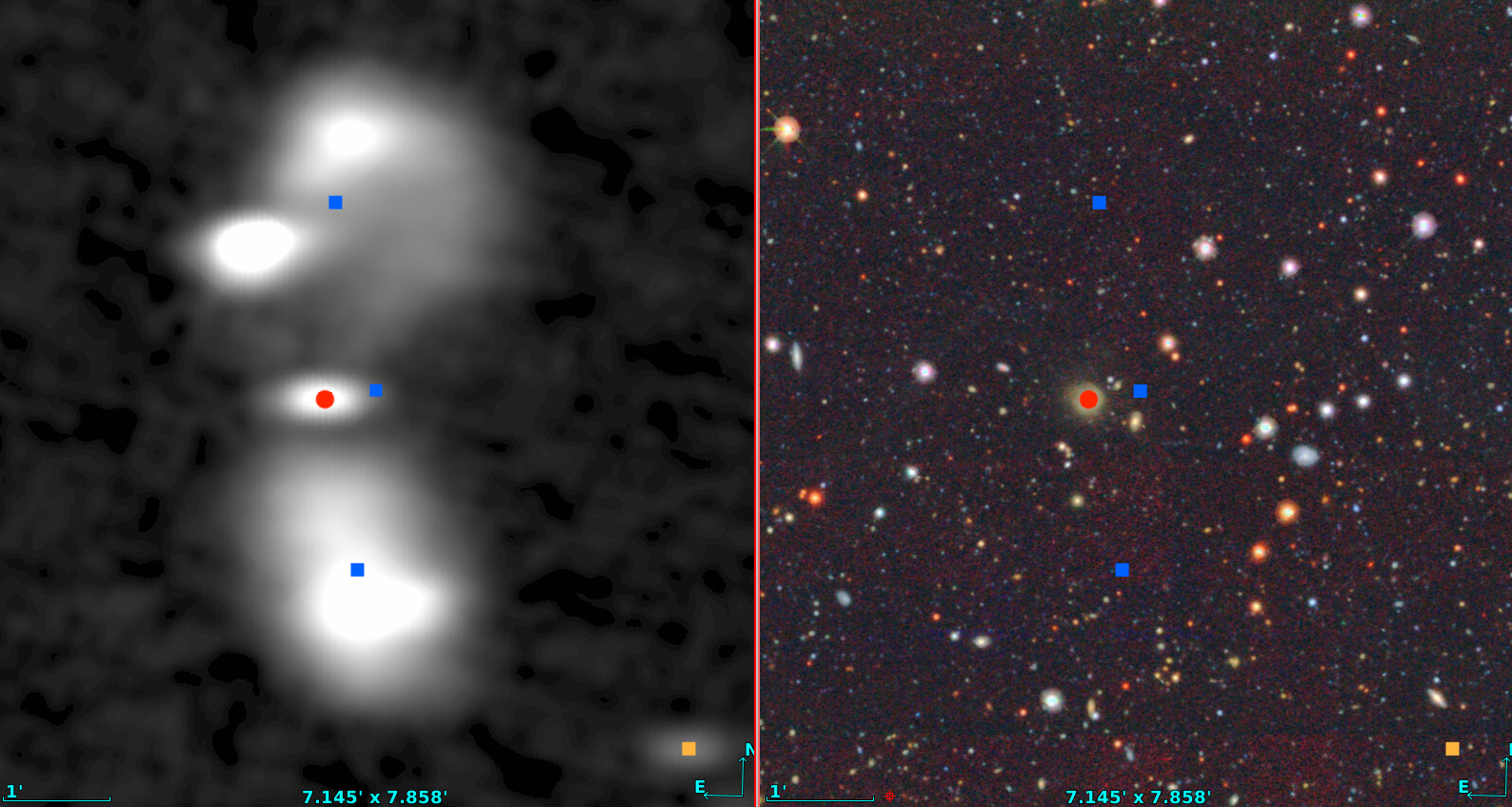}
\includegraphics[width=1\linewidth]{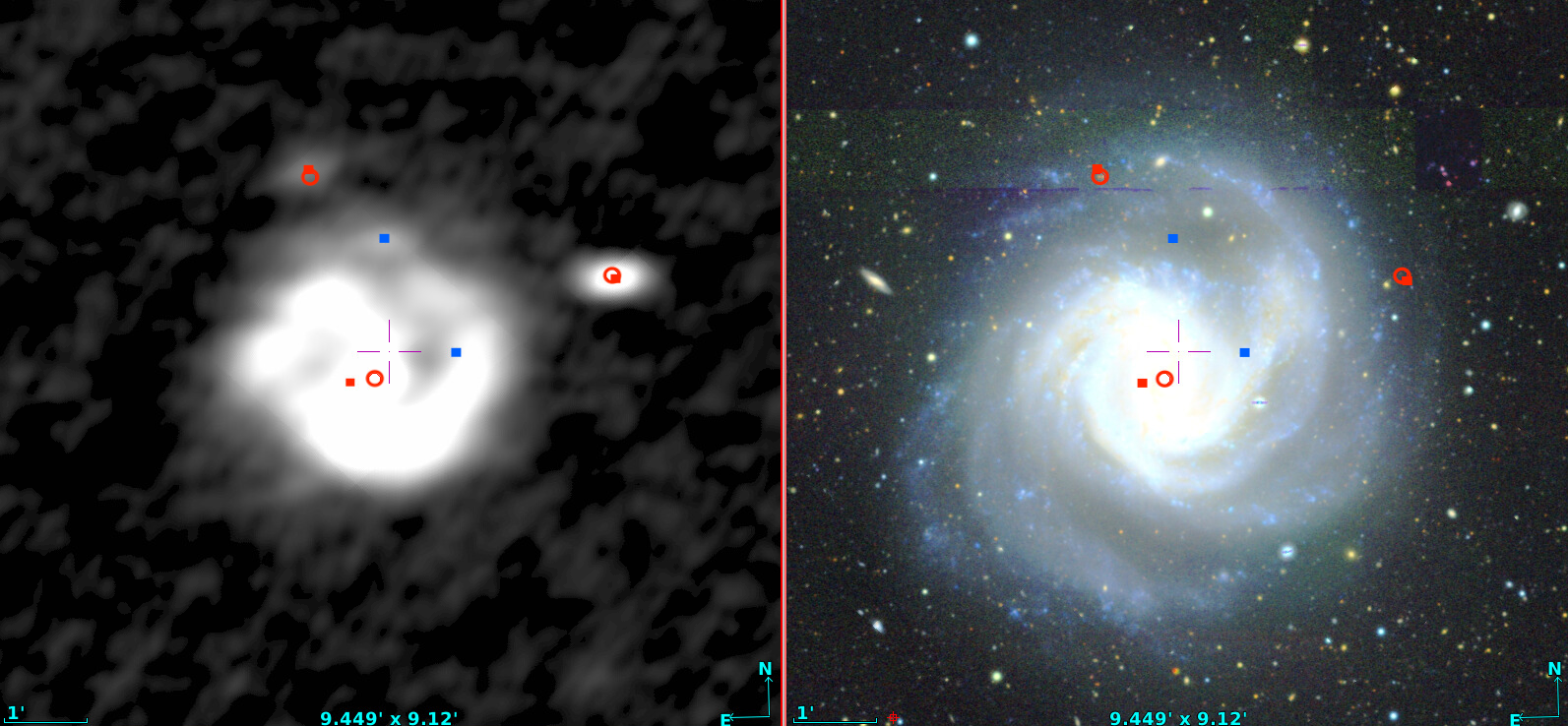}
\caption{Counterpart association and source merging in complex environments. Top row: 
Radio and optical images of a $7'\times7'$ region around MeerKLASS-UHF\_DR1\,J+101443.5-14421.1, a giant radio galaxy at $z=0.199$.  Sources are marked as in Fig.~\ref{fig:merging}.  The radio source is detected as four individual radio components and merged into one system (red), all matched to a single host galaxy. Bottom row: Radio and optical images of a $9'\times9'$ region around Messier 61 (MeerKLASS-UHF\_DR1\,J+122156.1+042823.5). The star-forming galaxy is split into three individual radio sources, while two background point sources are found in the vicinity, projected through the optical disk.}
\label{fig:merging2}
\end{center}
\vspace{-0.3cm}
\end{figure}

\subsection{Spectroscopic redshift assignment}
We use available spectroscopic redshifts either already provided by the KiDS and LS DR10 photometric catalogs or taken from public datasets such as SDSS DR18 \citep{SDSSDR18} and DESI DR1 \citep{DESIDR1}. We explicitly update existing SDSS redshifts with dedicated quasar-focused redshifts from SDSS DR16 \citep{SDSSQSO} to ensure robust redshifts for this source population. We further caution that spectroscopic redshifts from DESI DR1 at $1.6<z_{\mathrm{spec}}<1.7$ might be incorrect for a certain fraction of the sources. SDSS redshifts are used for sources in this redshift range if available. Remaining DESI DR1 sources in this redshift range have to be treated with caution, especially if they depart significantly from the photometric redshift measurement.

We match the list of known spectroscopic redshifts to the catalog of best optical hosts using a cross-match radius of $1''$.

\subsection{Treatment of complex radio sources}
Radio morphologies of extragalactic sources can be highly complex. In radio galaxies, one may detect emission from the lobes but not from the core and detect lobes and cores as separate radio sources.  Furthermore, orientation with respect to line of sight, Doppler boosting, and the radio galaxy environment can cause lobes to appear with different brightnesses and shapes. All these effects can lead to positional offsets between the radio detection and the host galaxy that significantly exceed the measurement uncertainties provided by the source-detection pipelines. This is part of the motivation for using empirically calibrated counterpart probabilities in SEDA instead of relying solely on nominal positional uncertainties. This approach compensates for modest radio-to-host offsets, for example those arising from resolved star-forming galaxies or from radio galaxies with small offsets between the lobes and the host. However, for larger lobe offsets the probability that a radio galaxy is split into at least two individual sources increases, as does the offset of each radio component from the true host. Such cases can result in radio sources without good optical or infrared counterparts at their nominal catalog positions.

To recover such systems, the SEDA algorithm is run 
in two passes. In the first pass, we search for the best counterpart using the nominal position given in the radio catalog. In the second pass, we select all radio sources without a good counterpart ($P_{\mathrm{true}}<0.5$) from the first pass and search for their nearest neighbors within this subset. We then calculate the midpoints between each of these radio sources and their nearest neighbor. In addition, we perform a source-detection run on the radio data with very low deblending settings and high detection thresholds to obtain intensity-weighted centers of extended sources that have been split into several components in the main imaging run and therefore appear as individual detections in the MeerKLASS catalogs. We then repeat the counterpart search using these additional centers for sources that did not have good counterparts in the first pass. 
The counterpart from the second pass is only assigned as the better counterpart if the difference in counterpart probability is $P_\mathrm{true,2nd}-P_{\mathrm{true,1st}}>0.2$ and $P_\mathrm{true,2nd}>0.5$. This limits noise biases due to multiple search attempts.

In Fig.~\ref{fig:merging}, we show a $4'\times4'$ region from the MeerKLASS UHF-band survey and the corresponding LS DR10 optical image. Four individual radio detections are highlighted as colored squares. In the first SEDA run, only the upper-right source (MeerKLASS-UHF\_DR1\,J+123954.5+034242.2) had a high-probability counterpart, because the clearly double-lobed radio galaxy was detected as a single radio source. In the second SEDA run, the search between neighboring radio sources revealed a high-probability counterpart (red circle) between the lobes of the lower-left source. The nearest radio source to the optical host, MeerKLASS-UHF\_DR1\,J+124001.6+034037.5, is assigned as the primary radio source, while the optical host is found to be a QSO at $z=1.88$. For the remaining radio source (yellow square), no high-probability counterpart was found.

\subsubsection{Flagging sources belonging to the same system}
In a final step, we attempt to flag radio sources that likely belong to the same host galaxy. Sources that have a good counterpart at their radio position form the basis of this procedure. These are typically star-forming galaxies, compact sources, and the cores of radio galaxies. For sources with a good counterpart in the second run, we check whether the identified host is already assigned to another radio source from the first run. If so, all corresponding radio sources are flagged as belonging to the same host. We add a column to the radio catalog that lists the source ID of the radio source closest to the host. We also combine the fluxes of the components and provide a column listing the total combined flux associated with the given host galaxy.

We then search among all remaining sources without good optical or infrared hosts for additional components of larger radio systems. This is done by using the original PyBDSF \citep{PyBDSF} output to connect components with islands, defined as connected pixels above a given detection threshold. Similarly, we use the output of the low-deblending source-detection run described above. If a source without a good counterpart shares an island with a source that has a good counterpart, we assign it to the same physical source, flag it as a component of another radio source, provide the corresponding source ID, and calculate the combined flux.

In Fig.~\ref{fig:merging2}, we show radio and optical images for two complex cases illustrating the challenges associated with flagging multiple radio components belonging to the same host galaxy. The upper case shows the giant radio galaxy associated with MeerKLASS-UHF\_DR1\,J+101443.5-14421.1 ($z=0.199$). The radio galaxy is detected as four individual radio sources, and the core is successfully assigned as the primary host. In the lower case, we show Messier 61 ($z=0.005$), which is resolved and split into three radio sources. The SEDA algorithm correctly merges these sources while keeping potential background sources separate.

\begin{figure}
\begin{center}
\centering
\includegraphics[width=1\linewidth]{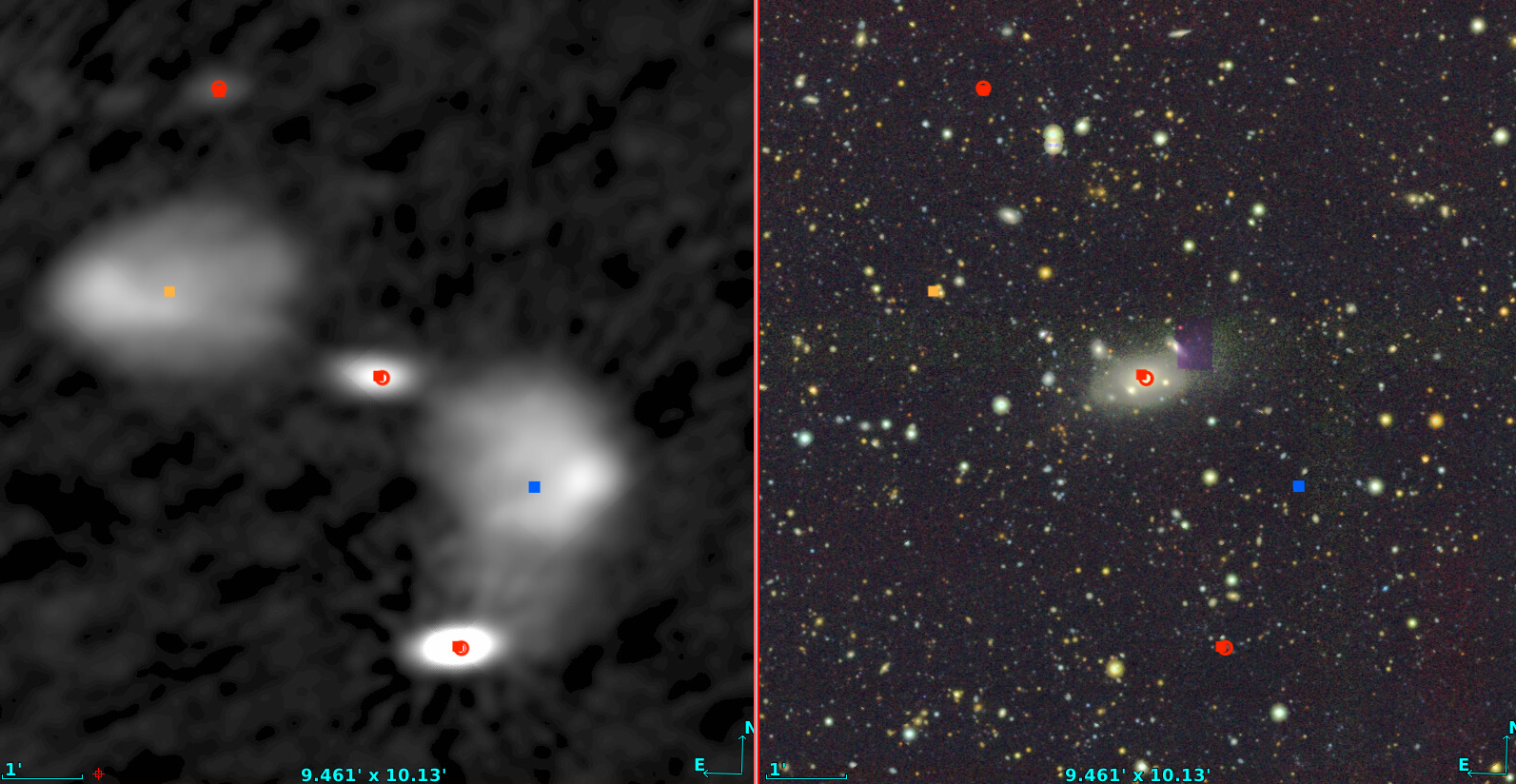}
\caption{Example of a failure mode in counterpart association and source merging. Similar to Figs.~\ref{fig:merging} and \ref{fig:merging2}, the panels show a $9'\times9'$ region centered on the giant radio galaxy associated with MeerKLASS-UHF\_DR1\,J+102407.6-020344.4. Both lobes (yellow and blue squares) are not merged with the core (central red square), because of their large separation (yellow lobe) and the projection of an unrelated source (blue lobe).}
\label{fig:merging3}
\end{center}
\vspace{-0.3cm}
\end{figure}

This procedure successfully assigns and merges multiple radio sources associated with the same physical host in many cases. However, a subset of systems remains where the automated process fails to merge components or assign the correct counterpart. This is particularly true for very extended, widely separated, or strongly asymmetric sources. One such example is shown in Fig.~\ref{fig:merging3} for the nearby ($z=0.052$) radio galaxy MeerKLASS-UHF\_DR1\,J+102407.6-020344.4. The $9'\times9'$ region centered on the core of the radio galaxy shows two large ($>2'$) radio lobes north-east and south-west of the core. The SEDA algorithm correctly associates the core with the host galaxy, while no connection to the core is found for the north-east lobe (yellow square), which remains listed as a source without a counterpart. The south-west lobe (blue square) is associated with the bright point source to the south (MeerKLASS-UHF\_DR1\,J+102403.6-020705.5). In this constellation, the total number of hosts is correct, but the lobe fluxes are not correctly assigned to the core, and in one case the flux is falsely assigned to an unrelated source. This type of failure mode depends strongly on redshift, because angular sizes and separations decrease with increasing redshift.

To test the performance for complex systems and source merging, we visually inspected several samples of potentially complex UHF-band sources. We constructed three subsets and inspected 50--60 sources from each. The first subset contains sources with the largest semi-major axes and signal-to-noise ratios (SNRs) greater than 100. The second subset contains sources with SNRs greater than 100 and total-to-peak flux ratios greater than 3. The third subset contains sources with SNRs greater than 100 and at least nine Gaussian components measured in the radio source catalog. These selections were designed to target the most challenging cases for the counterpart search.

Across these subsets, we find that in 62--69\% of cases our approach identifies the correct counterpart with $P_{\mathrm{true}}>0.5$. In 17--23\% of cases, the correct counterpart was not recovered and the associated and the best counterpart had $P_{\mathrm{true}}<0.5$, while in 10--15\% of cases an incorrect counterpart is assigned despite $P_{\mathrm{true}}>0.5$. The number of missed sources is approximately consistent with the overall expected incompleteness of the sample, discussed in the following subsection. The fraction of incorrectly assigned counterparts exceeds the expected $\sim5\%$ contamination because the inspected samples are intentionally biased toward difficult cases and because unrelated radio sources can be merged with a radio source that has a correct host assignment.

As a final visual test, we inspected 56 sources with $P_{\mathrm{true}}>0.5$ drawn from the $\sim1,300$ sources that are flagged as connected to another source but are not assigned as the primary host. We find that the counterpart is correctly assigned in 86\% of cases, that the source merging is incorrect in 12\% of cases, and that the optical counterpart is wrong in one case (2\%). The single wrong counterpart is consistent with the expected number of contaminants, while the 12\% fraction with incorrect source merging is slightly smaller than, but close to, the estimates for the more extreme subsets inspected above.

In summary, source merging works for 86\% of the inspected subset of non-primary sources associated with another source. For the most extreme and visually challenging subsets, the success rate and completeness of the counterpart catalog are mildly lower than average.

\subsection{Chance-association calibration of $P_{\mathrm{true}}$}

To calibrate the sample purity with respect to chance associations, we construct a control catalog by applying a uniform 6 arcmin offset to the MeerKLASS radio source positions and then rerunning the SEDA counterpart search. This position-displaced control catalog preserves the angular selection function and the clustering of the radio sources in the test positions, while breaking the physical association with their true optical or infrared hosts. It therefore provides an empirical estimate of the $P_{\mathrm{true}}$ distribution expected for chance associations.

We apply the same two-pass procedure to the control catalog as to the real radio catalog. In the second pass, which searches between neighboring radio sources, we restrict the number of control positions to match the number used for the real catalog. This ensures that the rate of chance associations is measured under the same number of search attempts. The resulting control measurement gives the number of associations expected by chance at a given $P_{\mathrm{true}}$ and can therefore be used to estimate the reliability of the counterpart catalog.

The $P_{\mathrm{true}}$ distribution measured for the real catalog is a mixture of true counterparts and unrelated associations. The shape of the unrelated-association component is estimated from the control catalog. Assuming that the low-$P_{\mathrm{true}}$ regime of the real catalog is dominated by unrelated associations, we rescale the control distribution such that it matches the real distribution at low $P_{\mathrm{true}}$. In the top panel of Fig.~9, we show the $P_{\mathrm{true}}$ distribution of UHF-band counterparts, together with the corresponding distribution from the control catalog and its rescaled version. The rescaled control distribution matches the observed distribution around real UHF-band sources well below $P_{\mathrm{true}}=0.2$. Repeating the same procedure for the L-band survey gives a consistent result.

Using the rescaled control distribution, we define the final calibrated association probability $P_{\mathrm{assoc}}$ as
\begin{equation} P_{\rm assoc}(P_{\rm true}) = 1 - \frac{ N_{\rm ctrl}^{\rm rescaled}(P_{\rm true}) }{ N_{\rm real}(P_{\rm true}) } . \label{eq:passoc} \end{equation}  
Here, $N_{\mathrm{real}}$ is the number of SEDA associations in the real catalog and $N^{\mathrm{rescaled}}_{\mathrm{ctrl}}$ is the rescaled number of associations in the control catalog.

\begin{figure}
\begin{center}
\centering
\includegraphics[width=1\linewidth]{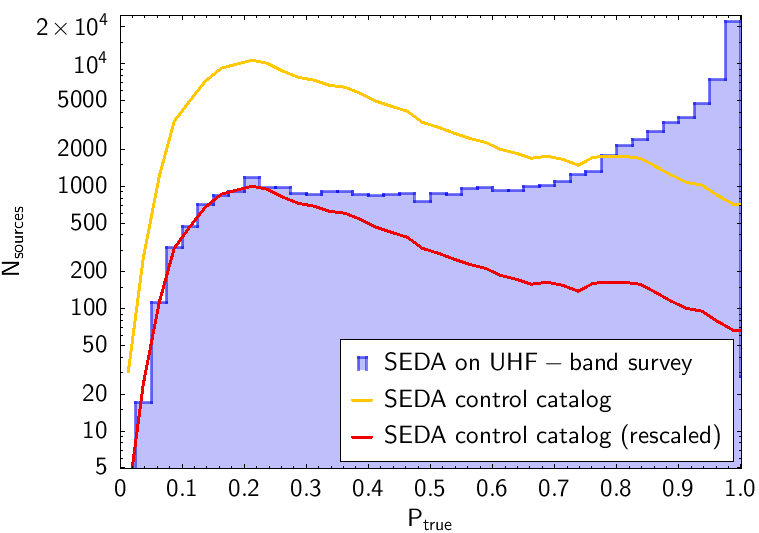}
\includegraphics[width=0.95\linewidth]{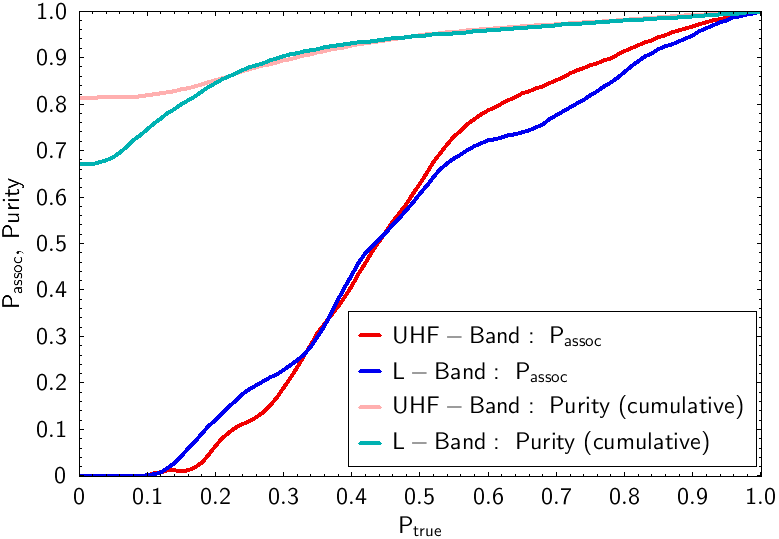}
\caption{Measurement of association probability $P_{\rm assoc}$ and sample purity. Top: Histogram of SEDA measurements of $P_{\mathrm{true}}$ for UHF-band catalog and on displaced positions as control sample. Bottom:  $P_{\rm assoc}$ given $P_{\mathrm{true}}$ for L-band and UHF-band catalogs and sample purity for samples selected by  $P_{\mathrm{true}}>P_{\mathrm{true, min}}$.}
\label{fig:ptruecalib}
\end{center}
\vspace{-0.3cm}
\end{figure}

The relation between $P_{\rm assoc}$ and $P_{\rm true}$ is shown for the UHF-band and L-band samples in the lower panel of Fig.~\ref{fig:ptruecalib}.  Over the whole range an increase in one estimator leads to an increase in the other. 
At $P_{\mathrm{true}}<0.2$, $P_{\mathrm{assoc}}$ approaches zero by construction because of the normalization of the control distribution. Around $P_{\mathrm{true}}\simeq0.4$, the two probability estimates become similar, while at higher $P_{\mathrm{true}}$ values the calibrated association probability exceeds the original SEDA estimate.

We also show in Fig.~\ref{fig:ptruecalib} the purity of the counterpart sample above a given threshold in $P_{\rm true}$ for both surveys. For thresholds above $P_{\rm true}=0.3$ the estimated purity appears to be very similar for both surveys, despite their differences in optical data, radio bands or depth. We find a purity of 94.7\% for $P_{\rm true}>0.5$ for both surveys, further increasing to 97\% for $P_{\rm true}>0.7$.
This relation allows users to select thresholds in either $P_{\mathrm{true}}$ or $P_{\mathrm{assoc}}$ depending on the desired balance between completeness and contamination.

\begin{figure}
\begin{center}
\centering
\includegraphics[width=1\linewidth]{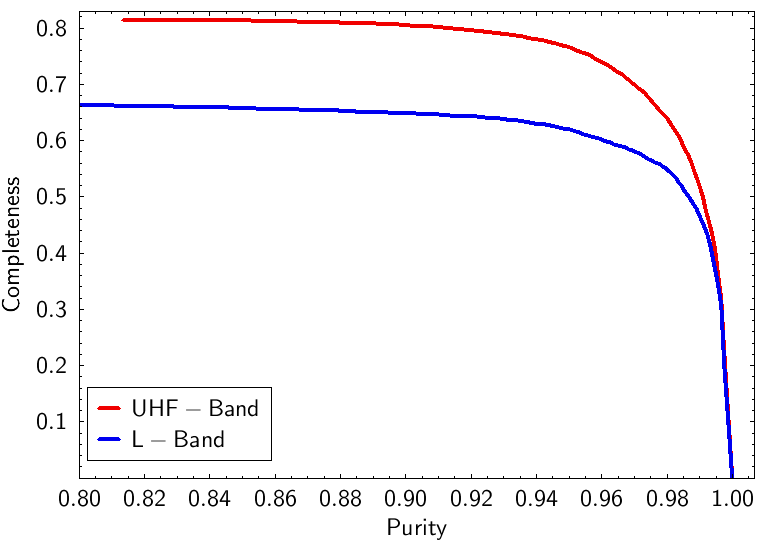}
\caption{Completeness versus purity for UHF and L-band surveys, assuming all radio detections are true sources.}
\label{fig:complpur}
\end{center}
\vspace{-0.3cm}
\end{figure}

In Fig.~\ref{fig:complpur}, we show completeness as a function of purity for the counterpart catalogs. For this estimate, we assume that all radio detections are real, while the number of true counterparts is estimated by summing $P_{\mathrm{assoc}}$ over the selected sample. The figure shows that purities well above 90\% can be reached with only a modest impact on completeness. We note that the completeness estimated in this way is mildly conservative: false detections in the radio catalog would lead to an overestimate of the number of true radio sources, while the normalization of the control distribution tends to underestimate the purity at very low $P_{\mathrm{true}}$.

\section{Survey specific differences}
\subsection{Application to the UHF-band survey}
The UHF-band footprint is well covered by LS DR10. In Fig.~\ref{fig:coverage}, we show the UHF-band radio image overlaid on the LS DR10 mask map. The dominant cause of masking is the proximity to bright stars, which can affect source detection and photometry. We recommend using the mask flags provided in the catalog, but we nevertheless provide counterpart candidates in masked regions to allow users to apply their own quality cuts.

The primary source of photometric redshifts is the photometric-redshift catalog provided by the Legacy Surveys team \citep{Zhou2023LRGPhotoz}. This catalog does not provide stellar masses, but includes a compilation of spectroscopic redshifts available at the time of its construction. We estimate stellar masses from the LS DR10 $z$-band magnitude and the $g-r$, $r-z$, and $z-W1$ colors, using galaxies with matched LS DR10 photometry and stellar masses from COSMOS \citep{Laigle} and GAMA \citep{Taylor2011GAMAStellarMass} as reference samples. 
The UHF-band footprint is also well covered by spectroscopic surveys, including SDSS \citep{SDSS}, GAMA, and DESI \citep{DESIDR1}, which leads to an enhanced fraction of counterparts with spectroscopic redshifts.

\subsection{Application to the L-band survey}
The L-band survey only has patchy coverage in LS DR10. The main optical dataset for counterpart identification is therefore KiDS DR5, whose coverage within the L-band footprint is shown in Fig.~\ref{fig:coverage}. In contrast to the LS DR10 coverage of the UHF footprint, KiDS DR5 does not cover the northern and southern edges of the L-band survey, and sources around bright stars are already excluded from the released catalog.

Source detection in KiDS DR5 is performed on deep $r$-band images, whereas LS DR10 uses combined $g,r,z$ detection images. While the KiDS $r$-band imaging is typically deeper than the corresponding LS DR10 band, the deep $z$-band data in LS DR10 are more efficient for detecting faint passive galaxies at high redshift. This creates a mild disadvantage for identifying high-redshift radio galaxies in KiDS alone, where even massive passive galaxies can become faint. Despite the inhomogeneous LS DR10 coverage in the L-band footprint, we therefore perform an additional LS DR10-based counterpart search.

For the L-band survey we provide two counterpart catalogs. The baseline catalog uses only KiDS DR5 counterparts and is restricted to the 87\% of the footprint covered by KiDS. The extended catalog is augmented with LS DR10 counterparts to provide larger and locally deeper coverage. In addition to increasing the available area, the LS DR10-based search allows us to compare KiDS- and LS DR10-based counterpart probabilities in the subset of the footprint with high-quality coverage in both datasets.

\begin{figure*}
\begin{center}
\centering
\includegraphics[width=0.475\linewidth]{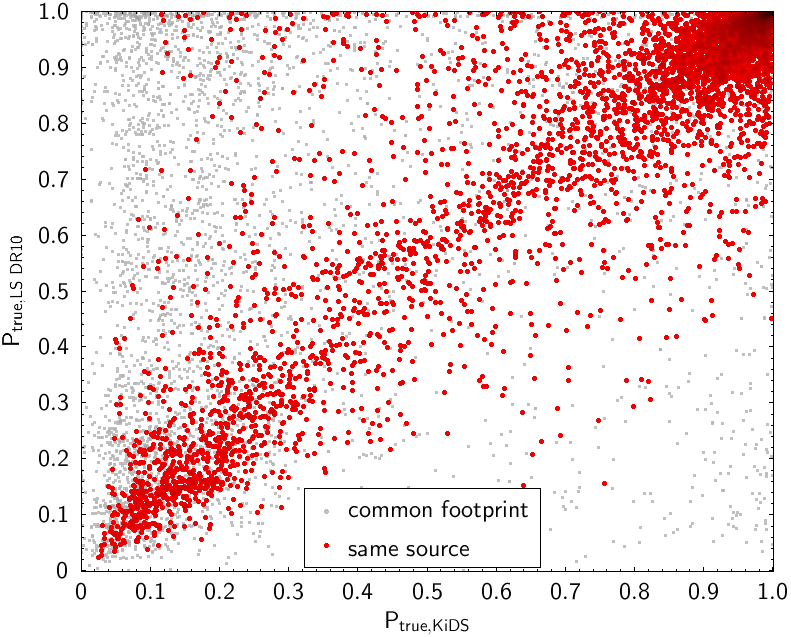}
\includegraphics[width=0.51\linewidth]{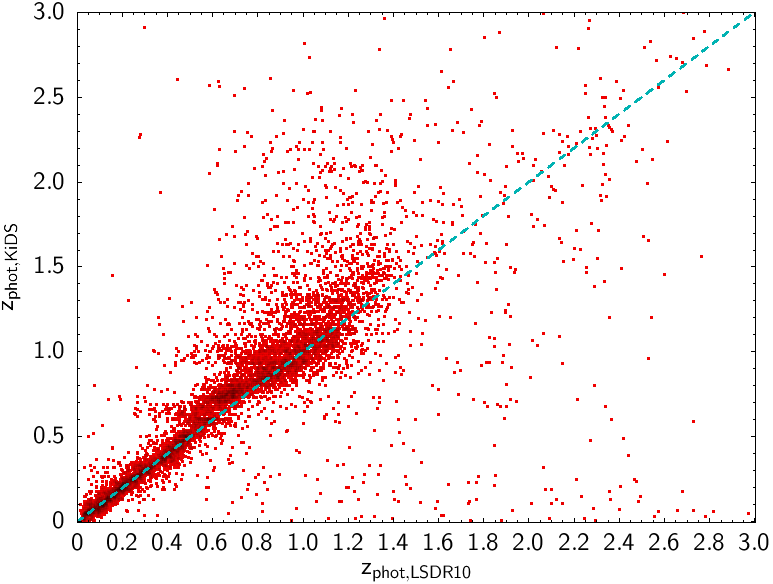}
\caption{Comparison of KiDS and LS DR10 results on L-band survey. Left: Comparison of L-band counterpart probabilities $P_{\rm true}$ for common counterparts in the joint high-quality KiDS and LS DR10 footprint (red). Sources with differing counterparts are shown in grey. Right: Comparison of photometric redshifts from KiDS and LS DR10 for common counterparts.}
\label{fig:ptruevsptrue}
\end{center}
\vspace{-0.3cm}
\end{figure*}

\subsection{Comparison between KiDS and LS DR10 data}
For the part of the L-band footprint with both KiDS DR5 and LS DR10 coverage, we compare the counterpart probabilities obtained from the two optical datasets. This comparison provides a useful consistency check on the SEDA probabilities and highlights survey-dependent selection effects. In general, high-probability counterparts are recovered consistently in both catalogs, especially for relatively bright and compact sources. Differences arise primarily for faint passive galaxies, sources close to the detection limit, and objects affected by masking or deblending differences between the two catalogs. The left panel of Fig.~\ref{fig:ptruevsptrue} illustrates the comparison of the KiDS- and LS DR10-based probabilities for matched counterparts in the common high-quality footprint. Sources with high $P_\mathrm{true}$ in only one of the two surveys are predominantly caused by differences in depth. This plot does not fully show sources with high $P_\mathrm{true}$ in both catalogs but with differing counterparts. Using a maximum matching distance of 2 arcsec and requiring $P_\mathrm{true}>0.5$, we find that 7.4\% of the sample has different counterparts. Increasing the threshold to $P_\mathrm{true}>0.9$ reduces the fraction with differing counterparts to 1.9\%. We visually inspected all $P_\mathrm{true}>0.9$ sources with differing counterparts and find that in 19\% of cases the KiDS counterpart is likely the correct counterpart, while in 28\% of cases the LS DR10 counterpart is preferred. In the remaining 53\% of cases, visual inspection does not clearly indicate which catalog provides the better counterpart. These ambiguous cases are typically caused by counterparts with small angular separations, making them equally plausible matches to the radio source.

In the right panel of Fig.~\ref{fig:ptruevsptrue}, we compare photometric redshifts of $P_\mathrm{true}>0.5$ counterparts. The majority of the $\sim8,000$ counterparts follow the one-to-one relation indicated by the cyan dashed line. The scatter increases at $z\gtrsim0.8$. Two additional features are apparent: a bias between the two photometric-redshift estimates at $0.5\leq z \leq 0.9$ and an apparent hard cut-off in the LS DR10 photo-z distribution at $z\sim1.5$. As outlined in Appendix~\ref{ap:photoz}, there is evidence that the bias at $0.5\leq z \leq 0.9$ is primarily caused by biased redshifts in the KiDS DR5 catalog. The apparent hard cut-off at $z\sim1.5$ is caused by biased photo-z estimates in LS DR10, likely driven by the lack of near-infrared photometry and the limited training sample for the photo-z calibration. The observed colors available in LS DR10 alone provide limited information for constraining the redshifts of passive galaxies beyond $z=1.5$, and the lack of passive galaxies at $z>1.5$ in the training sample causes galaxy photo-z estimates in LS DR10 to be limited to $z\approx1.5$. Only galaxies with quasar-like colors can reach higher photometric redshifts. A more detailed discussion of the redshift performance is provided in Appendix~\ref{ap:photoz}.

\begin{figure}
\begin{center}
\centering
\includegraphics[width=1\linewidth]{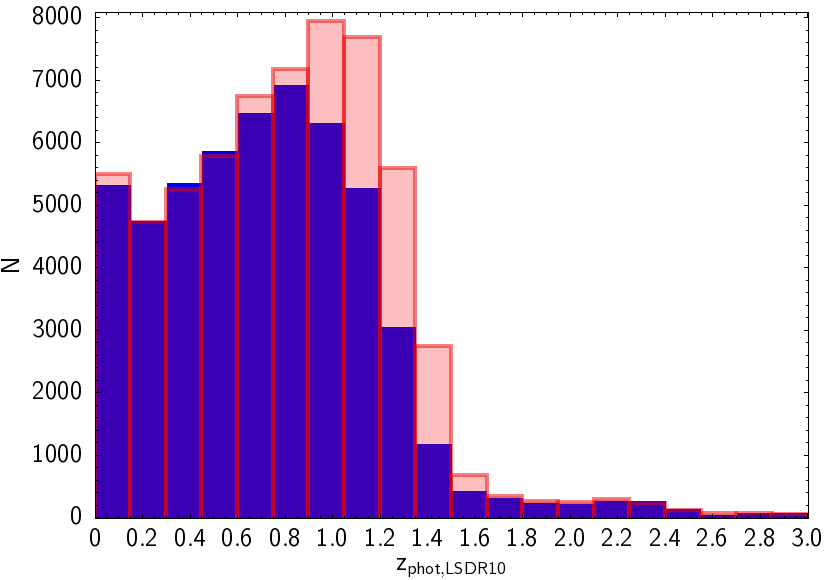}
\caption{Distribution of photometric redshifts of LS DR10 identified ($P_\mathrm{true}>0.5$) counterparts in the UHF-survey (red). Shown in blue is the distribution of counterparts also detected in KiDS DR5, scaled by a factor of 2.78 to account for partial coverage of the footprint.}
\label{fig:lsdr10kidsredshiftdist}
\end{center}
\vspace{-0.3cm}
\end{figure}

From Fig.~\ref{fig:complpur}, the completeness of the UHF-band counterparts reaches $\sim81\%$, whereas the L-band counterpart completeness reaches only $\sim66\%$. Besides possible differences caused by the radio data, one explanation may be the differences between the KiDS and LS DR10 datasets. While the LS DR10 data over the L-band survey area are very heterogeneous, KiDS covers $\sim36\%$ of the UHF-survey footprint. We therefore cross-match KiDS data with the LS DR10-based host galaxies to estimate the fraction of LS DR10 hosts that are also detected by KiDS. Differences between the two surveys should be driven predominantly by imaging depth and, at very low redshift, potentially by the different treatment of very extended nearby sources. In Fig.~\ref{fig:lsdr10kidsredshiftdist}, we show the redshift distribution of UHF-band hosts identified with LS DR10 data. We also show the distribution of the subpopulation also detected by KiDS DR5, scaled by a factor of 2.78 to account for the difference in survey coverage. The scale factor is derived by comparing the number of matches at $0.1<z_\mathrm{phot,LS DR10}<0.6$ with the total number of sources in this redshift range. Assuming that the KiDS data are complete in this redshift range, the ratio of all LS DR10 sources to KiDS-matched sources provides a direct estimate of the relative survey area covered by KiDS. The difference between the scaled KiDS-matched distribution and the full LS DR10 sample then provides an estimate of the KiDS completeness with respect to LS DR10, as well as the redshift range in which sources are missing from KiDS. We measure a KiDS completeness of $84\pm1\%$ relative to LS DR10, suggesting that this is the main reason for the lower counterpart fraction in the KiDS-based L-band follow-up. From Fig.~\ref{fig:lsdr10kidsredshiftdist}, one can see that the majority of the missing sources lie at $0.9<z<1.5$, the redshift regime dominated by radio galaxies with passive hosts. At higher redshifts, the relative incompleteness decreases because quasars are intrinsically brighter and bluer in the optical bands than passive galaxies.

In summary, the cross-comparison yields reasonable agreement in redshifts and $P_\mathrm{true}$ at $z<1$ and for bright sources. Photo-z estimates for passive hosts from LS DR10 are biased low for true redshifts $z>1.5$, while KiDS photo-z estimates show a mild bias at $0.5\leq z \leq 0.9$. The relative completeness of KiDS and LS DR10 is the main cause of the different counterpart fractions in the two radio surveys.

\begin{figure}
\begin{center}
\centering
\includegraphics[width=1\linewidth]{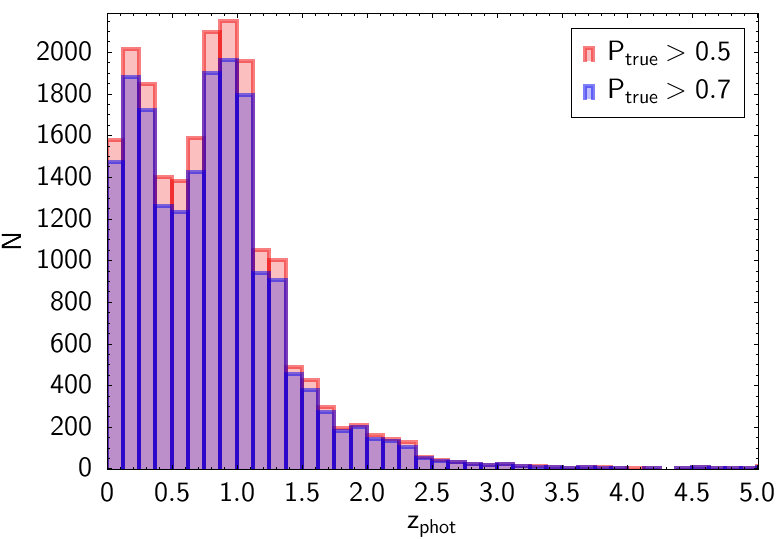}
\includegraphics[width=1\linewidth]{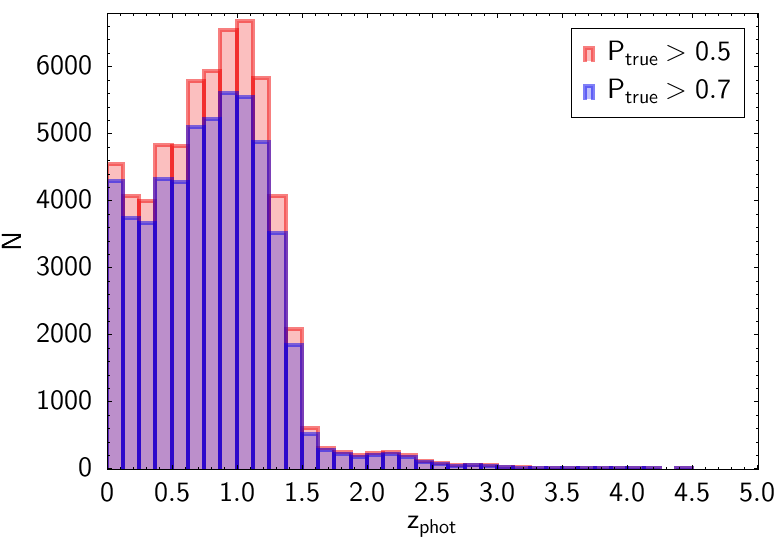}
\caption{Redshift distribution of confirmed counterparts in the L-band survey (top) and UHF-band survey (bottom).}
\label{fig:redshiftdist}
\end{center}
\vspace{-0.3cm}
\end{figure}

\section{Properties of the counterpart catalogs}\label{sec:results}
In this section we present the outcome of the counterpart search with SEDA for the L-band and UHF-band surveys. The counterpart catalogs will be provided on CDS after acceptance of the paper, the column desciption of the most relevant columns can be found in Appendix~\ref{ap:columndesc}. 
In this section focus on the overall counterpart fractions, the redshift distributions, and the comparison between photometric and spectroscopic redshifts. More detailed source-population studies, including the dependence on radio luminosity, stellar mass, and optical color, will be presented in future work.

\subsection{General properties of the resulting catalogs}
In Figs.~\ref{fig:ptruecalib} and~\ref{fig:complpur} we show key diagnostic quantities, including the sample purity as a function of the $P_\mathrm{true}$ threshold, the mapping between $P_\mathrm{true}$ and $P_\mathrm{assoc}$, and completeness as a function of purity. While users can define selection thresholds according to their scientific needs, we emphasize two useful reference values. The first is $P_\mathrm{assoc}=0.5$ ($P_\mathrm{true}\approx0.44$), which marks the point where the calibrated probability that a counterpart is a true association is 50\%. Including sources below this value adds candidates for which chance associations are more likely than true associations, which is likely undesirable for most scientific applications. The second threshold is $P_\mathrm{true}=0.5$ ($P_\mathrm{assoc}\approx0.61$), for which we find the estimated purity and completeness with respect to the maximum achievable completeness to be $\sim94.5\%$. This threshold therefore provides a balance between lost sources and number of chance associations remaining in the sample. In the subsequent discussion, we use $P_\mathrm{true}>0.5$ as the default threshold for showing results.

The main difference between the counterpart catalogs of the two surveys lies in their completeness. One possible explanation is related to differences in optical source detection and survey depth. Sources in KiDS are detected from deep $r$-band imaging data, while LS DR10 uses combined $g$, $r$, and $z$-band detection images. The typical hosts of radio galaxies are passive galaxies and therefore appear red in optical data. The inclusion of redder bands, such as the $z$ band, therefore gives LS DR10 an advantage for identifying hosts of high-redshift radio galaxies compared to the $r$-band-based KiDS detection. Additional contributions to the difference in completeness may come from the incomplete KiDS coverage of the L-band footprint and from differences in the depth of the radio surveys.

\subsubsection{L-band catalog}
We find 20,400 counterparts with $P_\mathrm{true}>0.5$ in the KiDS-based catalog, corresponding to 66\% of all L-band sources within the KiDS footprint. In the top panel of Fig.~\ref{fig:redshiftdist}, we show the redshift distribution of KiDS counterparts using the KiDS photometric redshifts for two probability thresholds, $P_\mathrm{true}>0.5$ and $P_\mathrm{true}>0.7$, corresponding to purities of approximately 95\% and 97\%, respectively. The redshift distribution shows three main features: a low-redshift peak at $z<0.4$, a second peak around $z\approx1$, and a steep drop-off followed by a long tail to high redshift.

These features broadly reflect the mixture of the main host populations. The low-redshift excess is associated with star-forming galaxies, while the peak around $z\approx1$ is dominated by radio galaxies. The subsequent steep decline is driven primarily by incompleteness in the optical catalog of host galaxies for radio galaxies. Cross-matched sources at $z>1.5$ are predominantly quasars, which remain detectable in optical and infrared data out to much higher redshifts than the typical radio AGN host galaxies.

We find spectroscopic redshifts for 22\% of the counterparts with $P_\mathrm{true}>0.5$. In the top panel of Fig.~\ref{fig:redshiftvsredshift}, we compare photometric and spectroscopic redshifts for these counterparts. The majority of sources show reasonable agreement, but we find biases at $0.6<z_\mathrm{phot}<0.8$ and a number of significant outliers. In particular, some high-redshift quasars are assigned low photometric redshifts by KiDS, illustrating the limitations of galaxy-optimized photometric-redshift estimates for rare quasar populations. 
We measure a median absolute deviation-based scatter of $\sigma_\mathrm{\Delta z/(1+z)}=0.03$ for $z_\mathrm{phot}<1$ and $\sigma_\mathrm{\Delta z/(1+z)}=0.14$ for $z_\mathrm{phot}>1.0$. The corresponding outlier fractions, defined as sources exceeding five times the measured scatter, are 6.5\% and 11.5\%, respectively. However, we note that the distribution of redshift differences, especially in the high-redshift bin, appears to consist of a subpopulation with well-estimated redshifts and a significantly broader component. Fitting a Gaussian function only to the well-estimated subpopulation would yield a much smaller scatter estimate and a much higher outlier fraction. The adopted measure is therefore intended to provide a robust summary statistic for comparison with the LS DR10 measurements for the UHF-band survey, rather than a full description of the observed distribution.

To increase coverage and completeness, we provide a merged catalog that combines counterparts from KiDS and LS DR10. LS DR10 counterparts are added to the catalog for all sources without KiDS coverage. In the overlapping region, we accept an LS DR10 counterpart if either $P_\mathrm{true, LSDR10}>0.7$ and $P_\mathrm{true, LSDR10}>P_\mathrm{true, KiDS}+0.1$, or if $P_\mathrm{true, LSDR10}>0.9$ and the LS DR10 host candidate is closer to the radio position than the KiDS host candidate. The first criterion limits the impact of adding false detections to the host-candidate list, while the second accounts for cases in which multiple radio sources were falsely associated with the same host because the true host was initially missed by KiDS. The merged catalog provides counterparts with $P_\mathrm{true}>0.5$ for 73\% of all candidates, adding a total of $\sim5000$ counterparts to the catalog. We note that, because of the patchy LS DR10 coverage over the L-band footprint, many of the LS DR10 sources do not have photo-z estimates. Consequently, the combined catalog lacks photo-z estimates for 14\% of the sources.

\subsubsection{UHF-band catalog}
Out of 75,823 UHF-band radio sources, we find 61,633 counterparts with $P_{\mathrm{true}}>0.5$, corresponding to 81\% of the sample. Increasing the threshold to $P_{\mathrm{true}}>0.7$ reduces the counterpart fraction to 71\%, while increasing the purity from approximately 95\% to 97\%. In the lower panel of Fig.~\ref{fig:redshiftdist}, we show the redshift distribution of the LS DR10 counterparts for the UHF-band radio sources. The distribution is qualitatively similar to that of the L-band sample, but the peak redshift is higher and the steep decline due to incompleteness in the optical catalogs begins at higher redshift than for the KiDS dataset. The median photometric redshift is $z=0.81$ for $P_{\mathrm{true}}>0.5$, compared to $z=0.77$ for the L-band sample selected with the same threshold.
For the LS DR10 counterparts, the decline appears steeper than for KiDS beyond $z\sim1.5$, potentially reflecting the lack of near-infrared photometric bands and the limited training data available to calibrate the machine-learning-based photometric-redshift code beyond this redshift. As a result, radio-galaxy hosts at $z\gtrsim1.5$ are often assigned photometric redshifts in the range $1<z_{\mathrm{phot}}<1.5$.

\begin{figure}
\begin{center}
\centering
\includegraphics[width=1\linewidth]{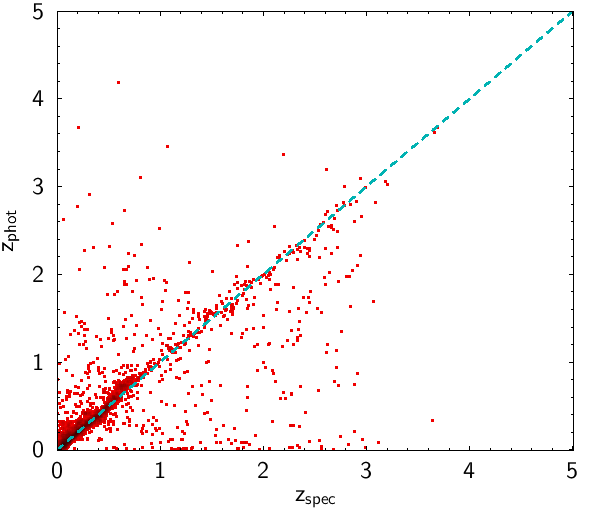}
\includegraphics[width=1\linewidth]{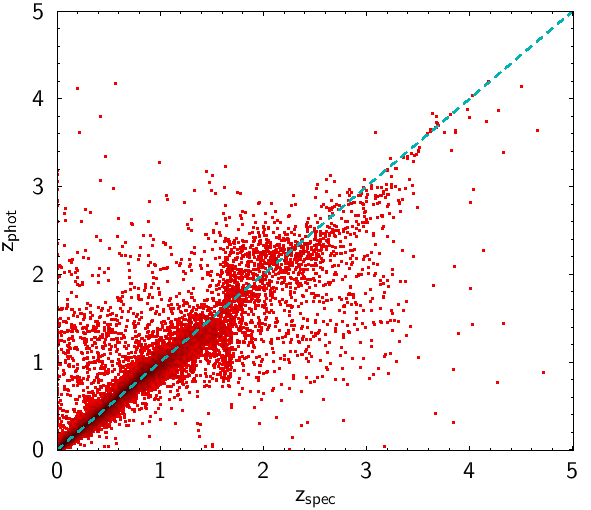}
\caption{Comparison between photometric and spectroscopic redshifts for counterparts in the L-band survey (top) and the UHF-band survey (bottom).}
\label{fig:redshiftvsredshift}
\end{center}
\vspace{-0.3cm}
\end{figure}

The SDSS and DESI coverage provides a high density of spectroscopic redshifts in the UHF-band footprint, resulting in spectroscopic redshifts for 44\% and 48\% of counterparts with $P_{\mathrm{true}}>0.5$ and $P_{\mathrm{true}}>0.7$, respectively. In Fig.~\ref{fig:redshiftvsredshift}, we compare spectroscopic and photometric redshifts for both counterpart catalogs. A zoom-in version showing the $z<1$ regime is provided in Fig.~\ref{fig:redshiftvsredshift2}. The photometric redshifts are largely unbiased for the bulk of the sample, but show substantial scatter at $z>1.5$. 
We also note a feature at $1.6<z_{\mathrm{spec}}<1.7$, which is a known issue related to the DESI spectroscopic sample. Spectroscopic redshifts in this redshift range should therefore be considered with care. We measure a median absolute deviation-based scatter of $\sigma_\mathrm{\Delta z/(1+z)}=0.02$ for $z_\mathrm{phot}<1$ and $\sigma_\mathrm{\Delta z/(1+z)}=0.12$ for $z_\mathrm{phot}>1.0$. The corresponding outlier fractions, defined as sources exceeding five times the measured scatter, are 4.5\% for the low-redshift bin and 6.7\% for the $z_\mathrm{phot}>1$ bin.

This mildly better performance compared to the KiDS-based counterparts of the L-band survey is less obvious in Fig.~\ref{fig:redshiftvsredshift}, because the UHF-band sample has six times more spectroscopic redshifts, leading to saturation effects close to the one-to-one relation.

\begin{figure}
\begin{center}
\centering
\includegraphics[width=1\linewidth]{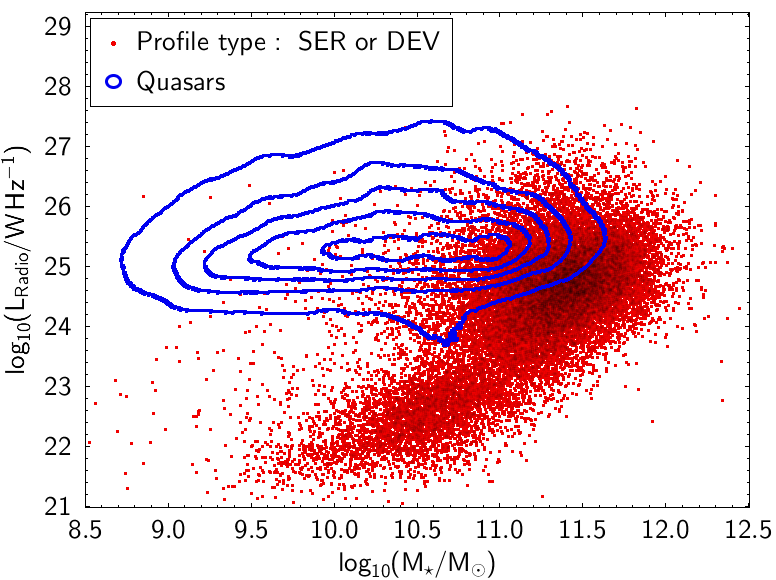}
\caption{Sources in the Radio luminosity-stellar mass plane from the UHF-band catalog: Shown in red are $\sim23,800$ sources with $P_\mathrm{true}>0.5$, $z_\mathrm{phot}<1$, and light profiles following S\'ersic (SER) or de~Vaucouleurs (DEV) profiles. Shown in blue are iso-density contours of sources following the WISE color selection for quasars. The contours are based on $\sim14,500$ sources over the full redshift range.}
\label{fig:lummass}
\end{center}
\vspace{-0.3cm}
\end{figure}

\subsection{Leveraging additional information from counterpart catalogs}
Besides photo-z and stellar mass, the counterpart catalogs offer additional information on host properties that can be used to study the samples in greater detail. We already use WISE colors and magnitudes to identify quasars within the counterpart candidates. Another potentially useful source of information is the morphology inferred from the optical imaging data. This information is provided in the LS DR10 catalog by the \texttt{TYPE} column, which gives the best-fitting light-profile model. Sources classified as S\'ersic (SER) or de~Vaucouleurs (DEV) are typically resolved sources in LS DR10. In the context of this work, this provides a simple way to select a comparatively clean sample of star-forming galaxies and radio galaxies without using color information from the catalog.

As an example, we show in Fig.~\ref{fig:lummass} galaxies with S\'ersic or de~Vaucouleurs profiles from the UHF counterpart catalog, selected with $P_\mathrm{true}>0.5$ and $z_\mathrm{phot}<1$. The counterparts selected in this way occupy a distinct region in the radio-luminosity--stellar-mass plane and follow a broad relation. The region with $\log_{10}(M_\star/M_\odot)<11$ and $\log L_\mathrm{Radio}<23.5$, where the stellar-mass--radio-luminosity relation is tighter, is occupied by star-forming galaxies, while radio galaxies occupy the high-mass, high-luminosity part of the distribution.
A clean quasar sample, in contrast, can be constructed by imposing WISE color cuts, such as those described in Sect.~\ref{sec:Quasars}, and excluding sources with S\'ersic or de~Vaucouleurs profiles. Quasars occupy the high-luminosity region of the radio-luminosity--stellar-mass plane and do not show a correlation between stellar mass and radio luminosity. For visibility, we show in Fig.~\ref{fig:lummass} only iso-density contours for the quasar population, because it overlaps with the population of radio galaxies.

This plot illustrates why stellar mass is not considered in the SEDA estimate of $P_\mathrm{true}$ for the quasar population, while it remains informative for the other host populations.
Future studies based on these counterpart catalogs can further use color or SED-fitting information, such as star-formation rates, to study the host populations and their properties in more detail.

\begin{figure*}
\begin{center}
\centering
\includegraphics[width=0.325\linewidth]{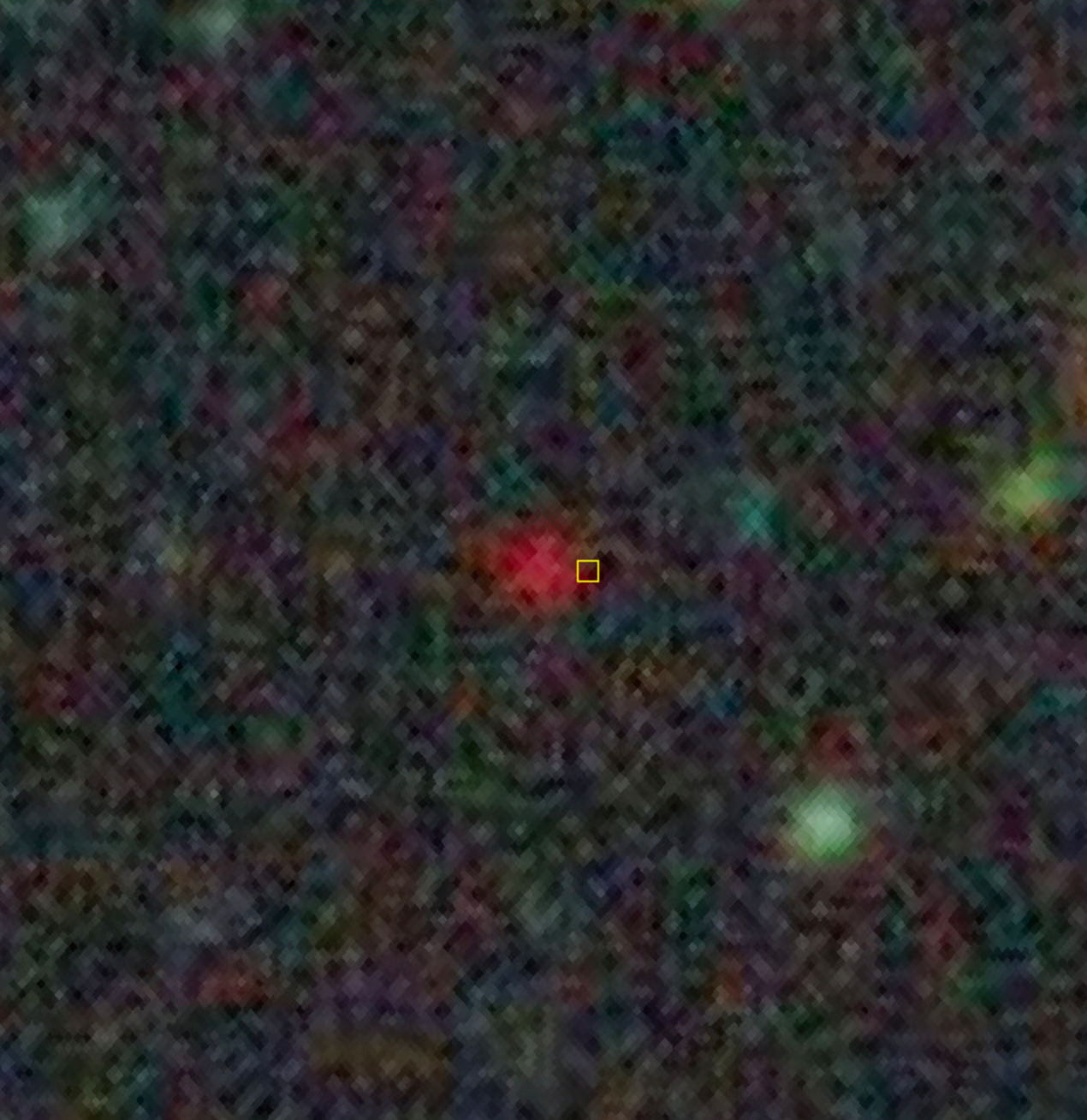}
\includegraphics[width=0.341\linewidth]{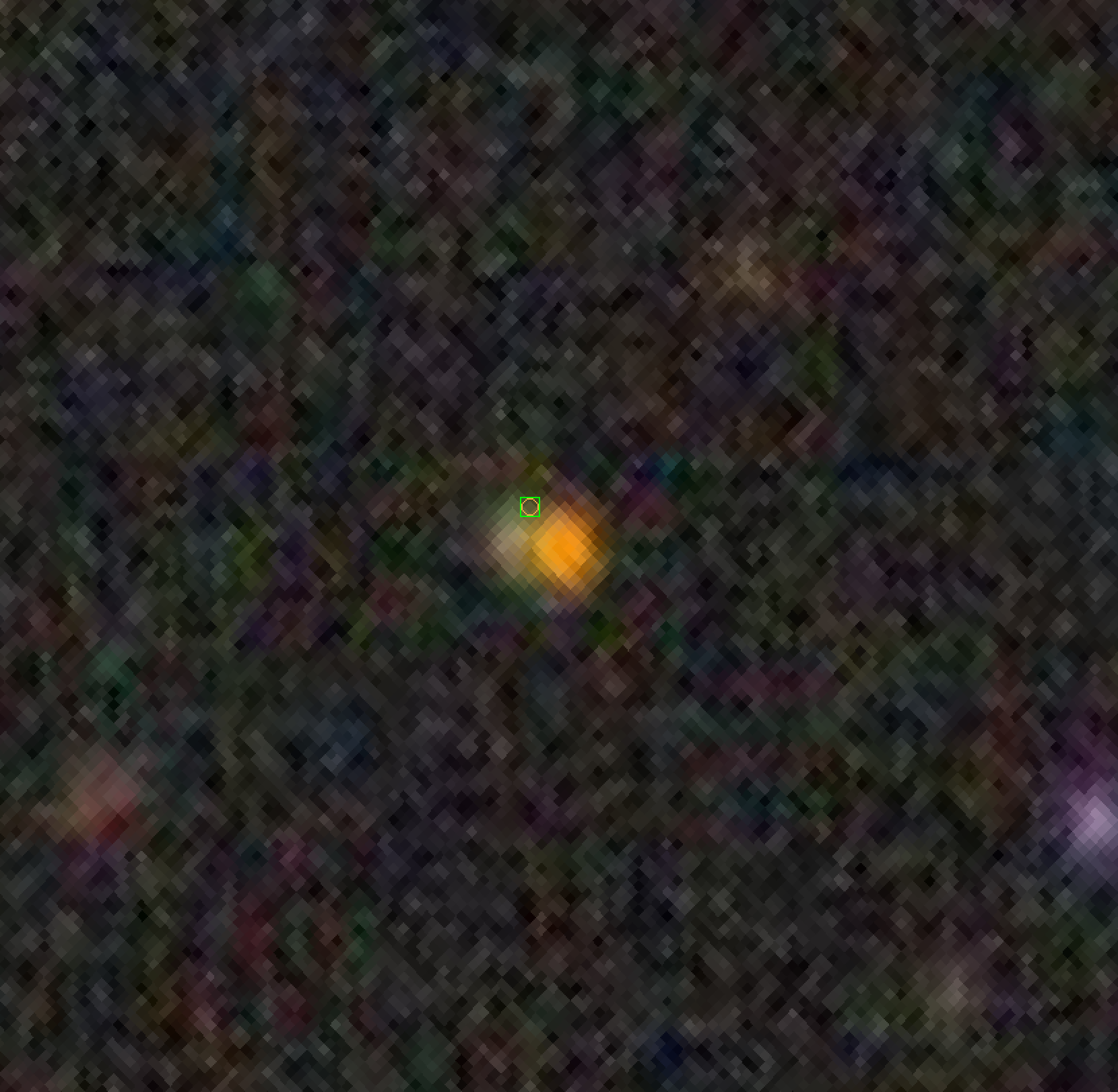}
\includegraphics[width=0.324\linewidth]{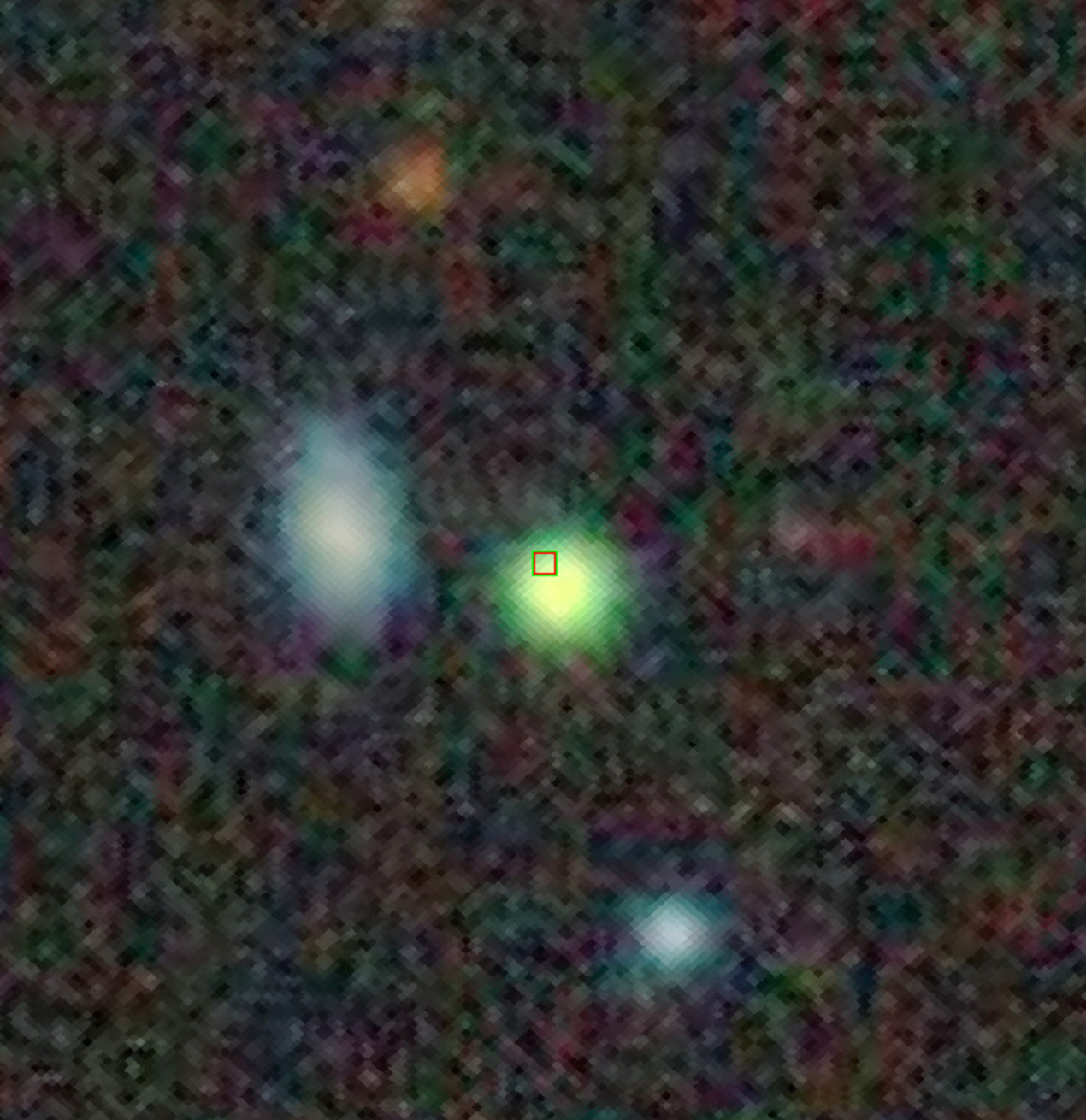}
\caption{High-redshift quasars identified in the L-band and UHF-band surveys. \emph{Left:} MeerKLASS\_L\_DR1\,J231818.3-311346.5 ($z=6.44$), the highest-redshift spectroscopically confirmed quasar identified in this work, located in the L-band footprint. \emph{Middle:} MeerKLASS-UHF\_DR1\,J+111111.8+053626.6 ($z=5.24$), the highest-redshift spectroscopically confirmed quasar in the UHF-band sample. \emph{Right:} MeerKLASS-UHF\_DR1\,J+102838.8-084437.8 ($z=4.27$), one of several quasars with $4<z_\mathrm{spec}<5$. Squares mark the position of the MeerKLASS detections.}
\label{fig:highzsources}
\end{center}
\vspace{-0.3cm}
\end{figure*}

\subsection{High-redshift sources}
High-redshift radio-loud quasars are rare objects, but they provide important probes of early black-hole growth, jet activity, and the build-up of massive galaxies at early cosmic times. Their identification in wide-area radio surveys is therefore an important by-product of large counterpart catalogs. Within both MeerKLASS surveys we find a tail of radio-loud quasars extending to high redshift.

At these redshifts, both photometric and spectroscopic redshifts require careful inspection. Photometric redshifts can fail for quasars, as shown in the top panel of Fig.~\ref{fig:redshiftvsredshift}, where KiDS assigns very low photometric redshifts to some spectroscopically confirmed high-redshift quasars. Spectroscopic redshifts can also fail if emission lines are misidentified. Redshifts for the highest-redshift candidates therefore need to be treated with care, especially when photometric and spectroscopic estimates are inconsistent.

Both optical catalogs used for host identification are based on source detection in optical bands. Given the imaging depth presented in Sects.~\ref{sec:Kids} and \ref{sec:lsdr10} and the typical stellar-mass distribution of radio-galaxy hosts, passive host galaxies are expected to become increasingly incomplete at $z\gtrsim1.4$. Quasar-mode hosts are intrinsically 
more luminous and bluer than passive galaxies and can therefore be detected to $z\approx4$ and beyond.

Quasar spectra show a strong suppression of flux blueward of Ly$\alpha$ at 1216~\AA\ due to absorption by intervening neutral hydrogen, together with an even sharper break below the Lyman limit at 912~\AA. At high redshift, these features move through the optical bands and produce the drop-out signatures widely used to identify high-redshift quasars \citep[e.g.][]{Fan2001}. For the present counterpart search, the Ly$\alpha$ absorption is the more relevant limitation for optical source detection. Ly$\alpha$ enters the $g$ band at $z\sim2.3$ and moves beyond it by $z\sim3.5$, leading to increasingly suppressed $g$-band flux and eventually to $g$-band drop-outs. It enters the $r$ band at $z\sim3.5$ and moves through it by $z\sim4.8$--5.0, making $r$-band-selected catalogs increasingly incomplete for quasars at $z\gtrsim4.5$--5. The Lyman-limit break enters the $r$ band only at higher redshift, around $z\sim5$, and reaches the red end of the $r$ band at $z\sim6.5$--6.7; however, by this point Ly$\alpha$ absorption has already strongly suppressed most of the $r$-band flux. While a dedicated high-redshift quasar search using drop-out selection is beyond the scope of this paper, these effects have direct consequences for the present counterpart search. The KiDS catalog used here is based on source detection in the $r$ band, implying that quasar identification becomes increasingly inefficient once Ly$\alpha$ absorption moves through the $r$ band and is expected to be strongly incomplete for quasars at $z\gtrsim5$.

Consequently, MeerKLASS\_L\_DR1\,J231818.3-311346.5, the counterpart with the highest spectroscopic redshift in our sample ($z=6.44$), was missed by KiDS but was identified by SEDA using LS DR10 within the MeerKLASS L-band footprint (see color composite image in Fig.~\ref{fig:highzsources}). MeerKLASS\_L\_DR1\,J231818.3-311346.5, also known as QSO~J2318-3113, was the most distant radio-loud quasar known at the time of its radio detection \citep{Ighina21} and remains one of the highest-redshift radio-loud quasars currently known. It is therefore encouraging that our automated counterpart search, which is optimized for the bulk of host galaxies, is capable of identifying counterparts even at the highest redshifts.

The source with the highest reliable spectroscopic redshift in the UHF-band sample is MeerKLASS-UHF\_DR1\,J+111111.8+053626.6, with $z=5.24$ \citep{Yang23}. In Fig.~\ref{fig:highzsources}, we show LS DR10 color-composite images of this source, QSO~J2318-3113, and a third quasar at $z=4.27$, illustrating the visual change in color as Ly$\alpha$ absorption and the Lyman break move through the optical bands.

While these results are encouraging, we expect a significant fraction of high-redshift quasars to lack spectroscopic redshifts or to have failed or imprecise photometric redshifts. We therefore plan a dedicated high-redshift-optimized study of radio-loud quasar candidates in the MeerKLASS fields.

\section{Conclusions}
We presented optical and infrared counterpart catalogs for radio sources detected in the MeerKLASS L-band and UHF-band surveys. The UHF-band catalog is based on LS DR10 imaging, while the L-band catalog primarily uses KiDS DR5 and is supplemented by LS DR10 where available. Together, these datasets provide wide-area optical and infrared coverage matched to the first MeerKLASS OTF continuum releases.

Counterparts were identified with the Stellar-mass Enhanced Density Association (SEDA) method, a data-driven framework that estimates empirical counterpart probabilities by comparing the distribution of candidate counterparts around radio positions to the corresponding distribution measured in a position-displaced control catalog.
The method uses radio-optical positional offset, stellar mass, and redshift to estimate the probability that a selected optical or infrared source is the true host of the radio emission.
Quasar-like sources are treated separately and identified with the help of WISE mid-infrared data.
The method therefore avoids imposing a strict separation between star-forming galaxies and radio AGN, while exploiting the fact that stellar mass is informative for counterpart identification in the non-quasar population.

For the L-band sample, we identify 20,400 KiDS counterparts with $P_\mathrm{true}>0.5$, corresponding to 66\% of the radio sources within the KiDS footprint. For the UHF-band sample, we identify 61,633 LS DR10 counterparts with $P_\mathrm{true}>0.5$, corresponding to 81\% of the radio sources. The spectroscopic-redshift fraction is 22\% for the L-band counterparts and 44\% for the UHF-band counterparts at this probability threshold, reflecting the stronger overlap of the UHF footprint with past and current spectroscopic surveys.

The resulting redshift distributions show the expected mixture of low-redshift star-forming galaxies, radio galaxies peaking around $z\approx1$, and a tail of high-redshift quasars. In particular, the SEDA algorithm enables the recovery of several radio-loud quasars at $z>4$, including QSO~J2318-3113 at $z=6.44$ in the L-band footprint and MeerKLASS-UHF\_DR1\,J+111111.8+053626.6 at $z=5.24$ in the UHF-band sample. These examples demonstrate that a counterpart search optimized for the bulk of the radio population can still recover rare high-redshift objects, although a dedicated high-redshift quasar search will be required for a complete census.

We also introduced an automated treatment of complex radio sources by repeating the counterpart search using additional candidate positions derived from neighboring radio components and radio source-detection runs with low-deblending. This approach allows many multi-component systems to be associated with a single optical host and provides combined radio fluxes for such systems. 
Very extended or strongly asymmetric sources remain challenging and will require additional algorithmic improvements in future catalog releases.

The catalogs presented here provide the basis for future studies of the MeerKLASS radio source population, including radio luminosity functions, the relative contributions of star formation and AGN activity, and searches for rare high-redshift radio quasars. 
The data used here represent only a small fraction of the full planned MeerKLASS survey, which is expected to cover 10,000\,deg$^2$.
The SEDA framework is designed to scale to future MeerKLASS data releases, where the larger sky coverage and greater number of radio sources will allow a more refined treatment of redshift, stellar mass, and source morphology.

\begin{acknowledgements}
    SM and JM acknowledge the support provided by the German Federal Ministry of Education and Research (BMBF) through the BMBF D-MeerKAT III award (number 05A23WM2). This funding was allocated via the `Verbundforschung' initiative. 
    SM also acknowledges support from the Excellence Cluster ORIGINS, which is funded by the Deutsche Forschungsgemeinschaft (DFG, German Research Foundation) under Germany's Excellence Strategy -- EXC-2094 -- 390783311. In addition, we acknowledge hardware support from the DFG-supported WAP program at LMU and thank the Rechenbetriebsgruppe within the Faculty of Physics.   
    SC acknowledges financial support from the South African National Research Foundation (Grant No. 84156) and the Inter-University Institute for Data Intensive Astronomy (IDIA).
    The radio data used in this study are available in the SARAO Online Archive \href{https://archive.sarao.ac.za}{(https://archive.sarao.ac.za)} with proposal ID SCI-20210212-MS-01 (L-Band) and ID SCI-20220822-MS-01 (UHF-band).

    We gratefully acknowledge the NSF supported Legacy Survey program consisting of three major individual and complementary projects: the Dark Energy Camera Legacy Survey (DECaLS; Proposal ID \#2014B-0404; PIs: David Schlegel and Arjun Dey), the Beijing-Arizona Sky Survey (BASS; NOAO Prop. ID \#2015A-0801; PIs: Zhou Xu and Xiaohui Fan), and the Mayall z-band Legacy Survey (MzLS; Prop. ID \#2016A-0453; PI: Arjun Dey). See also \href{https://www.legacysurvey.org/acknowledgment} for detailed acknowledgements.
    The Photometric Redshifts for the Legacy Surveys (PRLS) catalog used in this paper was produced thanks to funding from the U.S. Department of Energy Office of Science, Office of High Energy Physics via grant DE-SC0007914.

Based on KiDS data obtained from the ESO Science Archive Facility with DOI: https://doi.org/10.18727/archive/37, and https://doi.eso.org/10.18727/archive/59 and on data products produced by the KiDS consortium.
This research has made use of ``Aladin sky atlas'' developed at CDS, Strasbourg Observatory, France \citep{Aladin} and \textsc{topcat} \citep{Topcat}.
\end{acknowledgements}

%


\bibliography{myRefs.bib}

\begin{appendix}

\section{Column description of the counterpart catalogs}\label{ap:columndesc}
In Tables~\ref{tab:UHFcolumns} and~\ref{tab:Lcolumns}, we provide an overview of the main columns added by the counterpart search to the original MeerKLASS catalog. The column names are similar for the L-band and UHF-band catalogs, except for columns specific to the KiDS and LS DR10 photometric catalogs. The combined KiDS and LS DR10 catalog follows the same naming convention and contains additional columns with names ending in \texttt{COMB}, which merge entries from both catalogs. We note that this merged catalog is heterogeneous and should be treated with care. No attempt was made to merge multiple radio components derived from a mix of KiDS and LS DR10 counterparts.

Additional SEDA-based catalog entries, such as the nearest counterpart or the best counterpart before component merging, can be provided upon request and follow a similar naming convention.

\begin{table}[]
    \centering
    \caption{SEDA-based columns added to the original MeerKLASS columns for the main UHF-band counterpart catalog.}
    \label{tab:UHFcolumns}

\begin{tabular}{ l l }
  \multicolumn{1}{c}{Name} &
  \multicolumn{1}{c}{Description}  \\
\hline \hline
  RA\_MEERKLASS & Right ascension MeerKLASS\\
  E\_RA\_MEERKLASS & error on Right ascension\\
  DEC\_MEERKLASS & Declination MeerKLASS\\
  E\_DEC\_MEERKLASS & error on Declination\\
  RA\_HOST\_FINAL & Right ascension counterpart\\
  DEC\_HOST\_FINAL & Declination counterpart \\
  TYPE\_FINAL & LS DR10 Morph. model\\
  STAR\_GAL\_FINAL & optical star galaxy separator\\
  G\_MAG\_FINAL & g-band magnitude\\
  R\_MAG\_FINAL & r-band magnitude\\
  I\_MAG\_FINAL & i-band magnitude\\
  Z\_MAG\_FINAL & z-band magnitude\\
  W1\_MAG\_FINAL & w1-band magnitude\\
  W2\_MAG\_FINAL & w2-band magnitude\\
  G\_MAG\_E\_FINAL & error on g-band magnitude\\
  R\_MAG\_E\_FINAL & error on r-band magnitude\\
  I\_MAG\_E\_FINAL & error on i-band magnitude\\
  Z\_MAG\_E\_FINAL & error on z-band magnitude\\
  W1\_MAG\_E\_FINAL & error on w1-band magnitude\\
  W2\_MAG\_E\_FINAL & error on w2-band magnitude\\
  Z\_SPEC\_FINAL & spec-z\\
  Z\_PHOT\_MEDIAN\_FINAL & robust photo-z (default)\\
  Z\_PHOT\_STD\_FINAL & uncertainty on photo-z\\
  LG\_MSTAR\_FINAL & stellar mass\\
  P\_OFFSET\_FINAL & prob. solely using offset\\
  ELL\_OFFSET\_FINAL & radio-optical offset\\
  P\_TRUE\_FINAL & $P_\mathrm{true}$\\
  TOTAL\_FLUX\_COMB & radio flux incl. \\
  & other components\\
  HOST\_FLAG & 0 if component of other \\
  &radio source\\
  SOURCE\_ID\_HOST & source ID if only component \\
  P\_ASSOC\_FINAL & $P_\mathrm{assoc}$\\
  Purity\_CUMULATIVE & total cumulative purity \\
\hline\end{tabular}
\end{table}

\begin{table}[]
    \centering
    \caption{SEDA-based columns added to the original MeerKLASS columns for the main L-band counterpart catalog.}
    \label{tab:Lcolumns}
\begin{tabular}{ l l  }
  \multicolumn{1}{c}{Name} &
  \multicolumn{1}{c}{Description}  \\
\hline \hline
  RA\_MEERKLASS & Right ascension MeerKLASS \\
  E\_RA\_MEERKLASS & error on Right ascension \\
  DEC\_MEERKLASS & Declination MeerKLASS \\
  E\_DEC\_MEERKLASS & error on Declination\\
  RA\_HOST\_FINAL & Right ascension counterpart\\
  DEC\_HOST\_FINAL & Declination counterpart \\
  MAG\_GAAP\_G\_FINAL & g-band magnitude\\
  MAG\_GAAP\_R\_FINAL & r-band magnitude\\
  MAG\_GAAP\_I1\_FINAL & i-band magnitude\\
  MAG\_GAAP\_Z\_FINAL & z-band magnitude\\
  W1\_MAG\_FINAL & w1-band magnitude\\
  W2\_MAG\_FINAL & w2-band magnitude\\
  MAGERR\_GAAP\_G\_FINAL & error on g-band magnitude\\
  MAGERR\_GAAP\_R\_FINAL & error on r-band magnitude\\
  MAGERR\_GAAP\_I1\_FINAL & error on i-band magnitude\\
  MAGERR\_GAAP\_Z\_FINAL & error on z-band magnitude\\
  W1\_MAG\_E\_FINAL & error on w1-band magnitude\\
  W2\_MAG\_E\_FINAL & error on w2-band magnitude\\
  Z\_SPEC\_FINAL & spec-z\\
  Z\_BC\_FINAL & best photo-z\\
  Z\_BC\_ERR\_FINAL & error on photo-z\\
  MSTAR\_BESTFIT\_FINAL & stellar mass\\
  P\_OFFSET\_FINAL & prob. solely using offset\\
  ELL\_OFFSET\_FINAL & radio-optical offset\\
    P\_TRUE\_FINAL & $P_\mathrm{true}$\\
  TOTAL\_FLUX\_COMB & radio flux incl. \\
  & other components\\
  HOST\_FLAG & 0 if component of other \\
  &radio source\\
  SOURCE\_ID\_HOST & source ID if only component \\
  P\_ASSOC\_FINAL & $P_\mathrm{assoc}$\\
  Purity\_CUMULATIVE & total cumulative purity \\
\hline
\end{tabular}
\end{table}

\section{Photo-z performance}\label{ap:photoz}
\begin{figure}
\begin{center}
\centering
\includegraphics[width=1\linewidth]{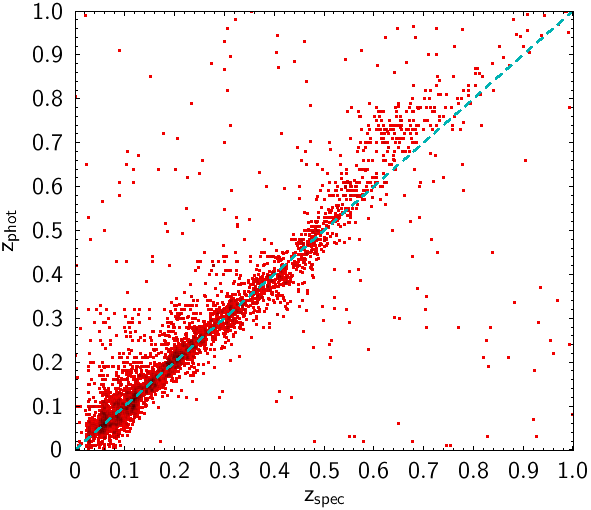}
\includegraphics[width=1\linewidth]{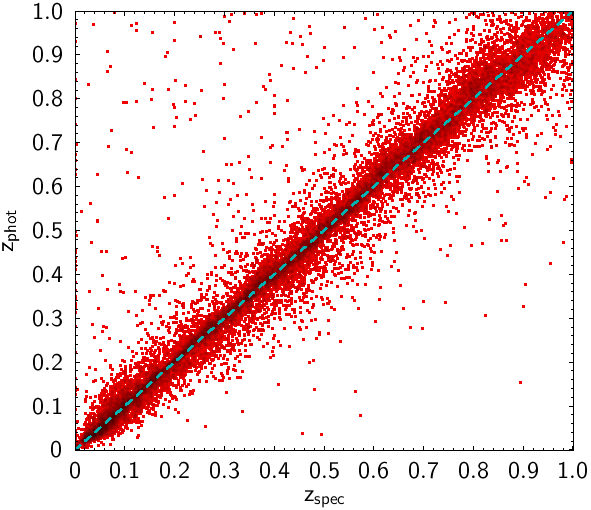}
\caption{Redshift comparison of confirmed counterparts in the L-band survey (top) and the UHF-band survey (bottom).}
\label{fig:redshiftvsredshift2}
\end{center}
\vspace{-0.3cm}
\end{figure}

In Fig.~\ref{fig:redshiftvsredshift2}, we show photometric redshifts plotted against spectroscopic measurements out to $z=1$ for the L-band (KiDS) and UHF-band (LS DR10) counterparts. Although KiDS uses ten photometric bands, in contrast to the five bands used in LS DR10, we see a bias of $\Delta z\approx0.05$ at $z\approx0.65$ for the host galaxies in our sample. The redshifts based on LS DR10, shown in the bottom panel of Fig.~\ref{fig:redshiftvsredshift2}, do not show a comparable bias. The visual impression of broader photo-z scatter in the LS DR10 sample is misleading, because the number of sources populating the two plots differs by more than a factor of five. With $\sigma_\mathrm{\Delta z/(1+z)}=0.02$ for $z<1$ and an outlier fraction of 4.5\%, the performance of LS DR10 is in fact better than that of KiDS, for which we find $\sigma_\mathrm{\Delta z/(1+z)}=0.03$ and an outlier fraction of 6.5\%.

\begin{figure}
\begin{center}
\centering
\includegraphics[width=1\linewidth]{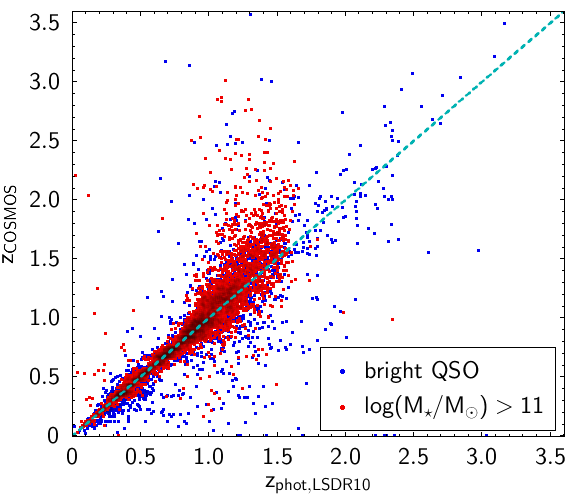}
\caption{Comparison of LS DR10 photo-z estimates and COSMOS photometric and spectroscopic redshift measurements. COSMOS galaxies with stellar masses similar to those of radio galaxies ($\log(M_\star/M_\odot)>11$) are shown in red. Galaxies following the bright QSO selection applied in the SEDA runs on MeerKLASS candidates are shown in blue.}
\label{fig:redshifscosmos}
\end{center}
\vspace{-0.3cm}
\end{figure}

To improve our understanding of the LS DR10 photo-z performance and to test whether features seen in the right panel of Fig.~\ref{fig:ptruevsptrue} are caused by LS DR10 measurements, we use the overlap between LS DR10 and the COSMOS photo-z catalog \citep{2022ApJS..258...11W}. After cross-matching sources within 1.5 arcsec, we select two subsamples representing two of the three main host types of MeerKLASS sources: quasars and massive galaxies. We expect star-forming galaxies to lie mostly at low redshift, where photo-z estimates perform well. For the quasar sample, we apply the same color cuts as used for the bright quasar sample defined in SEDA, while for massive galaxies we impose a stellar-mass cut of $\log(M_\star/M_\odot)>11$ in COSMOS to select galaxies similar to the typical hosts of classical radio galaxies in our sample. In Fig.~\ref{fig:redshifscosmos}, we show the comparison between LS DR10 photo-z estimates and high-quality COSMOS photometric redshifts.

The observed behavior appears similar to that seen in Fig.~\ref{fig:ptruevsptrue}. While quasar-like sources extend beyond $z\approx1.5$, the subset of massive galaxies appears to be truncated at $z_\mathrm{phot,LS DR10}\approx1.5$. The redshift distribution of the same galaxies measured in COSMOS extends out to $z\sim3$. There are two main reasons for this behavior in the LS DR10 redshifts. First, the available bands and corresponding colors provide limited information on redshifts beyond $z\approx1.6$. Second, the calibration of the redshift measurements themselves is based on spectroscopic redshifts and bright photometric galaxies. Because of the lack of redshifts for massive non-quasar galaxies that are faint in the optical bands, the applied machine-learning technique is not able to correctly predict redshifts beyond $z\approx1.5$, as the corresponding galaxies are not represented in the training sample.

Future studies that rely on photometric redshifts of massive and passive galaxies beyond $z=1$ should take this limitation of LS DR10 photo-z estimates into account.

\FloatBarrier 
\clearpage

\end{appendix}
\end{document}